\documentclass[10pt,aps,twocolumn,prb,longbibliography,showpacs,floatfix,superscriptaddress]{revtex4-2}

\usepackage{placeins}
\usepackage{graphicx,color}
\usepackage{amsfonts}
\usepackage[figuresright]{rotating}  
\usepackage{amssymb}
\usepackage{bm}
\usepackage{amsmath}
\usepackage{mathtools}
\usepackage[colorinlistoftodos,prependcaption,textsize=tiny]{todonotes}
\usepackage{xargs}                       
\usepackage{psfrag}
\usepackage{multirow}
\usepackage{tabularx}
\usepackage{textcomp}
\usepackage{units}
\usepackage{lipsum} 
\usepackage{soul}
\usepackage{titlesec}
\usepackage{times}
\usepackage{enumitem}
\usepackage{xcolor}
\usepackage{csquotes}
\usepackage{hyperref}

\DeclareMathAlphabet\mathbfcal{OMS}{cmsy}{b}{n}
\graphicspath{Images}
\titlespacing*{\section}
{0pt}{1ex plus .25ex}{1ex plus 1ex}
\titlespacing*{\subsection}
{0pt}{1ex plus .25ex}{1ex plus 1ex}
\titlespacing*{\subsubsection}
{0pt}{1ex plus .25ex}{1ex plus 1ex}

\titleformat{\section}
 {\fontsize{10}{15}\bfseries }
  {\thesection}
  {1em}
  {}

\titleformat{\subsection}
  {\bfseries }
  {\thesubsection}
  {2em}
  {}

\titleformat{\subsubsection}
  {\itshape}
  {\thesubsubsection}
  {2em}
  {}

\def\beq{\begin{eqnarray}}
\def\eeq{\end{eqnarray}}
\def\c{\hspace{2pt}}

\newcommand{\ket}[1]{\left| #1 \right>} 
\newcommand{\bra}[1]{\left< #1 \right|} 
\let\baraccent=\= 
\renewcommand{\=}[1]{\stackrel{#1}{=}} 
\newcommand{\bk}{\boldsymbol{k}} 
\newcommand{\mc}[1]{\mathcal{ #1}} 

\makeatletter

\titleclass{\subsubsubsection}{straight}[\subsection]

\newcounter{subsubsubsection}[subsubsection]
\renewcommand\thesubsubsubsection{\thesubsubsection.\arabic{subsubsubsection}}

\titleformat{\subsubsubsection}
   {\normalfont\normalsize\itshape \centering}{\thesubsubsubsection}{1em}{}
\titlespacing*{\subsubsubsection}
{0pt}{1ex plus .3ex}{1ex plus .3ex}

\makeatletter
\renewcommand\paragraph{\@startsection{paragraph}{5}{\z@}%
  {3.25ex \@plus1ex \@minus.2ex}%
  {-1em}%
  {\normalfont\normalsize}}
\renewcommand\subparagraph{\@startsection{subparagraph}{6}{\parindent}%
  {3.25ex \@plus1ex \@minus .2ex}%
  {-1em}%
  {\normalfont\normalsize}}
\def\toclevel@subsubsubsection{4}
\def\toclevel@paragraph{5}
\def\toclevel@paragraph{6}
\def\l@subsubsubsection{\@dottedtocline{4}{7em}{4em}}
\def\l@paragraph{\@dottedtocline{5}{10em}{5em}}
\def\l@subparagraph{\@dottedtocline{6}{14em}{6em}}
\makeatother

\newcommandx{\unsure}[2][1=]{\todo[linecolor=red,backgroundcolor=red!25,bordercolor=red,#1]{#2}}
\newcommandx{\change}[2][1=]{\todo[linecolor=blue,backgroundcolor=blue!25,bordercolor=blue,#1]{#2}}
\newcommandx{\info}[2][1=]{\todo[linecolor=green,backgroundcolor=green!25,bordercolor=green,#1]{#2}}
\newcommandx{\improvement}[2][1=]{\todo[linecolor=yellow,backgroundcolor=yellow!25,bordercolor=yellow,#1]{#2}}
\newcommandx{\thiswillnotshow}[2][1=]{\todo[disable,#1]{#2}}
\usepackage[english]{babel}

\begin{document}

\title{Microscopic field theories of the quantum skyrmion Hall effect}
\author{Vinay Patil}
\affiliation{Max Planck Institute for Chemical Physics of Solids, Nöthnitzer Strasse 40, 01187 Dresden, Germany}
\affiliation{Max Planck Institute for the Physics of Complex Systems, Nöthnitzer Strasse 38, 01187 Dresden, Germany}

\author{Archi Banerjee}
\affiliation{Max Planck Institute for Chemical Physics of Solids, Nöthnitzer Strasse 40, 01187 Dresden, Germany}
\affiliation{Max Planck Institute for the Physics of Complex Systems, Nöthnitzer Strasse 38, 01187 Dresden, Germany}
\affiliation{SUPA, School of Physics and Astronomy, University of St Andrews, St Andrews KY16 9SS, United Kingdom}

\author{Ashley M. Cook}
\affiliation{Max Planck Institute for Chemical Physics of Solids, Nöthnitzer Strasse 40, 01187 Dresden, Germany}
\affiliation{Max Planck Institute for the Physics of Complex Systems, Nöthnitzer Strasse 38, 01187 Dresden, Germany}

\begin{abstract}
     We construct effective field theories of the quantum skyrmion Hall effect from matrix Chern-Simons theory for $N$ electrons, corresponding to matrix dimension $N$. We first consider a quantum Hall droplet within finite $N$ matrix Chern-Simons theory. Taking into account the differential geometry of the matrix Chern-Simons droplet for a partially-filled fuzzy two-sphere, we first generalize the quantization procedure by replacing the Poisson bracket, a classical Lie derivative, with a quantum counterpart, the Lie derivative for a deformed fuzzy sphere. This yields the topological invariant introduced and applied in earlier works on the quantum skyrmion Hall effect, corresponding to previously unidentified fusion rules  at finite and even small $N$. This is consistent with treatment of a spin $S$ of multiplicity $2S+1$ as a quantum Hall droplet within matrix Chern-Simons theory for $N=2S+1$ spinless electrons and a generalization of a Jain composite particle for a Laughlin state. We then construct $D$-dimensional arrays of potentially coupled small $N$ matrix Chern-Simons droplets as effective field theories of the quantum skyrmion Hall effect. In higher-symmetry constructions, this yields what appears to be a D+1 dimensional $U(N)$ Yang-Mills theory---but actually contains $\delta$ extra fuzzy dimensions from the finite $N$ MCS theory as well as deformations from $U(N)$ due to partial filling of the fuzzy spheres. In this construction, the Chern-Simons level is $k+1$ for each small $N$ droplet, while the entire array can be interpreted as an unbounded matrix Chern-Simons theory at level $k$. Such constructions at $k=2$ are consistent with earlier numerical results for the multiplicative Chern insulator of the quantum skyrmion Hall effect. We also formulate the quantum skyrmion Hall effect in terms of a Lagrangian for an array of coupled, potentially distinct, small $N$ droplets within anisotropic fuzzification. We discuss the relevance of these results to spin lattice models and lattice gauge theories.
\end{abstract}
\maketitle
\tableofcontents
\section{Introduction}

Research into many branches of modern physics is now evolving at a rapid pace, buoyed by considerable technological advances over the past few decades. However, progress on some of the most fundamental questions of its development remains stalled. Notably, challenges identified roughly a half-century ago in condensed matter and high-energy physics remain, such that the following comments by Keimer~\emph{et al.}~\cite{keimer2015quantum} focused on one of these issues, namely the mechanisms of high-temperature superconductivity, are largely still relevant~\cite{zhou2021high, craig2022snowmass}. 

\enquote{Originally inspired by the desire to find out why superconductivity can
happen at a high temperature, condensed matter scientists engaged in a
relentless effort to unravel the physics of copper oxides. \textelp{} The bottom line
is that the existing theoretical machinery seems inadequate to describe both
the rich physics of the pseudogap phase and the nature of the strange
metal phase. \textelp{} In another development,
the practitioners of quantum information and string theory have landed
in the same territory, finding to their surprise that they are struggling with many of the same issues as condensed matter physicists.}

That such issues could range across such a broad expanse of modern physics suggests they stem from fundamental, potentially long-standing problem(s). The present work cannot hope to tackle these issues directly. However, it does address a long-held assumption concerning treatment of spin in quantum mechanics recently-identified in discovery of the quantum skyrmion Hall effect (QSkHE)~\cite{qskhe, cook2022multiplicative, cook2023topological, PhysRevB.108.045144, ay2024signatures, winter2025observable, banerjee2024multiplicative}. The QSkHE is a generalization of the framework of the quantum Hall effect (QHE), deep-rooted enough to impact all of these areas of research as the discovery that a spin degree of freedom (DOF) of a quantum system, which has previously been treated as an internal DOF, or isospin~\cite{pauli_uber_1925,Pauli1927ZurQD}, can encode $\delta>0$ number of spatial dimensions affecting the topology of a system with this spin DOF. A system with $D$ thermodynamically-large spatial dimensions plus these $\delta$ dimensions from spin can realize topologically non-trivial phases of matter of intrinsic dimensionality $D+\delta+1$, the $+1$ being the time dimension, with signatures distinct from those of $D+1$ dimensional topological states. Based on results of the present work, the generalization can also be stated as follows: a single quantum spin $S$ with $S$ small is more accurately modeled as an almost point-like quantum Hall droplet with $2S+1$ electrons, with previously-unidentified phenomena of quantum Hall states including bulk-boundary correspondence, quasi-particle/hole excitations (which are the simplest notion of quantum skyrmions of the QSkHE), fractional statistics, incompressibility, and topological ground-state degeneracy. This result is consistent with recent efforts in the high-energy literature to encode spatial dimensions in spin DOFs for $N$ large or infinite\cite{aschieri2007, aschieri2004dimensional, Aschieri:2004vh, Chatzistavrakidis:2010tq}, but reveals treatment of spin as an internal DOF---effectively just a label for quantum states---is an uncontrolled approximation dating back to introduction of quantum mechanics~\cite{1925ZPhy...33..879H, 1925ZPhy...34..858B, 1926ZPhy...35..557B}. 

While the QSkHE is motivated by gradual accumulation of observations of a large body of phenomena reported in a variety of lattice models~\cite{ay2024signatures,banerjee2024multiplicative, qskhe, cook2022multiplicative, cook2023topological, PhysRevB.108.045144, ay2024signatures, winter2025observable, banerjee2024multiplicative} though with some potential signatures in past experiments identified~\cite{ma2015unexpected}, these hard-won insights are increasingly understood from first principles derivations. In particular, recent work identifies effective field theories (EFTs) of the QSkHE consistent with these phenomena, finding they are well understood within the framework of extra fuzzy dimensions applied in the regime of small $N$, while also introducing more general formulations~\cite{qskhe,patil2024effective}.

The present work deepens understanding of these phenomena from first principles by introducing microscopic field theories of the QSkHE. Motivated by topological response signatures of one class of topological phases within the framework of the QSkHE, the multiplicative topological phases introduced by Cook and Moore~\cite{cook2022multiplicative}, microscopic field theories of the QSkHE are constructed as generalizations of matrix Chern-Simons (MCS) theory of Susskind for finite matrix dimension $N$ as developed primarily by Polychronakos and others~\cite{polychronakos2001quantum, polychronakos2001quantumcyl, morariu_fractional_2005, Cappelli:2009pn}. However, the present work identifies the quantization procedure first applied by Susskind and later employed by Polychronakos as potentially inconsistent with differential geometry of the fuzzy spaces over which MCS is defined. The generalized quantization procedure discussed here instead takes into account deformations of partially-filled fuzzy spaces identified as an area of future work by Polychronakos\cite{morariu_fractional_2005} and mathematical formulation of fuzzy spheres by Madore~\cite{Madore:1991bw}, to derive the topological invariant previously introduced to characterize the QSkHE~\cite{ay2024signatures, banerjee2024multiplicative, patil2024effective} from first principles. As a consequence, the present work demonstrates that a single quantum spin $S$, of multiplicity $2S+1$, is more accurately minimally treated within the framework of the QSkHE as a $2+1$ dimensional ($2+1$ D) MCS theory for a quantum Hall (QH) droplet of $2S+1$ electrons, or equivalently a $1+1$ D Calogero model, specifically a Calogero-Sutherland model (CSM), of $2S+1$ electrons. We refer to such a QH droplet within MCS theory/CSM as an MCS/CSM droplet. More general microscopic theories of the QSkHE are then constructed as $D$-dimensional arrays of multiple of such MCS theories or CSMs, which are coupled to one another with each individually encoding $2S+1$ electrons with $S$ potentially small. Given the fundamental link between the CSM and many other topics central to modern theoretical physics~\cite{minahan1994interacting,Gorsky:1993pe, ha_fractional_1995, kawakami_novel_1993,Yue_1998,yu_microscopic_1999, PhysRevB.43.11025, Tong:2016kpv, langmann_elliptic_2025, sakamoto_correspondence_2005, berntson_conformal_2025}, connecting the QSkHE to the CSM illustrates how microscopic theories of the QSkHE are relevant to these topics as well as significant in treatments of spin lattice models and generalizations to quantum dimer models and lattice gauge theories~\cite{PhysRevLett.61.2376}.

The manuscript is organized into four sections as follows: in the next part of this section, we review matrix Chern-Simons (MCS) theory, the Calogero-Sutherland model (CSM), the quantum skyrmion Hall effect (QSkHE), and related phenomenon of the QSkHE focusing on the multiplicative Chern insulator (MCI), which are the primary subjects in this work. Section~\ref{cquant} is devoted to study of a single MCS/CSM droplet, and then pair of MCS/CSM droplets, for cases of small $N$. At this level, we generalize quantization constraints for MCS theory due to partial filling of the non-commutative matrix space by electrons. We relate the corresponding discussion to the details of the topological invariant identified in previous works on the QSkHE, and modeling of individual unit cells of the lattice systems realizing signatures and phenomena of the QSkHE. Section~\ref{tiling} and section~\ref{array} instead focus on constructing arrays of $M$ such MCS/CSM droplets, each droplet with $N$ small, while the number of these MCS/CSM droplets, $M$, is greater or much greater than $N$. In section~\ref{tiling}, we explore the correspondence between the QSkHE and Yang Mills theory by constructing arrays from many copies of a small $N$ MCS/CSM droplet. Section~\ref{array} is then devoted to construction of lower-symmetry arrays of multiple MCS/CSM droplets, each with a small number of electrons relative to the number of droplets in the array, to describe the QSkHE e.g., for cases of disorder. To do so, we expand on the concept of anisotropic fuzzification introduced in previous work~\cite{patil2024effective}, to realize counterparts of the arrays in section~\ref{array}, in which each MCS/CSM droplet with a small number of electrons may be distinct.

\subsection{Brief review of matrix Chern-Simons theory}\label{secrew}
In this review section, we introduce Susskind's matrix Chern-Simons (MCS) theory at Chern-Simons (CS) level $k$ for an infinite number of electrons~\cite{susskind2001quantum}. We then introduce the MCS Lagrangian of Polychronakos for a finite number of electrons~\cite{polychronakos2001quantum, polychronakos2001quantumcyl, morariu_fractional_2005, Cappelli:2009pn}, and summarize his arguments for relating the MCS to the CSM for quantization of the filling fraction. The equivalence of MCS formalism with the Laughlin wavefunctions is discussed in~\cite{hellerman2001quantum} for finite and $N\to\infty$ limit, by showing one-to-one correspondence between the wavefunctions and constructing isomorphisms between the respective Hilbert spaces.

In derivation of MCS theory, Susskind~\cite{susskind2001quantum, Bahcall:1991an} discusses the problem of $N$ electrons on a plane subjected to an out-of-plane magnetic field of strength $B$, for which the most essential terms of the $N$ particle Lagrangian are
\begin{equation}
    L=\frac{B}{2}\sum_{i=1}^N\epsilon^{ab}\dot{x}^i_ax^i_b.
\end{equation}
He presents an action for a corresponding charged fluid with average density $\rho_0$ and position coordinates $\{x_a(\vec{y})\}$, with $\vec{y}$ being co-moving co-ordinates \cite{polychronakos2007non}, in a two-dimensional plane. Subjecting the charged fluid to a magnetic field of strength $B$ yields
\begin{align}
L' &= \frac{e B \rho_0}{2} \epsilon_{ab} \int d^2 y \left[\left(\dot{x}_a - \frac{1}{2 \pi \rho_0} \{x_a, A_0 \} \right)x_b + \frac{\epsilon_{ab}}{2\pi \rho_0}A_0\right],
\end{align}
at which point the Poisson bracket notation is introduced as
\begin{align}\label{poisson}
    \{F(y), G(y) \} = \epsilon_{ij} \partial_i F \partial_j G.
\end{align}
A counterpart discretized theory for $N$ spinless electrons, in which $N \times N$ matrices $\{X_a\}$ replace the fluid coordinates $\{x_a\}$, can be derived by different approaches~\cite{susskind2001quantum}, which yield the same outcome as the action~\cite{polychronakos2001quantum}
\begin{align}
S&= \int dt \frac{B}{2} \epsilon_{ab}\text{Tr}\left(\dot{X}_a - i \left[ A_0, X_a \right] \right)X_b + 2 \theta A_0.
\label{SusskindMatrixModel}
\end{align}
Here, Tr is the matrix trace over the Hilbert space and $\left[f,g \right]$ denotes a commutator of matrices $f$ and $g$. $\theta$ is a non-commutativity parameter defining the area in real-space occupied by an electron. 

The equation of motion (EOM) obtained by varying $A_0$ in the action $S$ and setting the variation in the action with respect to variation in $A_0$ to zero, which is referred to as the \textit{Gauss law} throughout the present work, takes the form
\begin{equation}
    [X_1,X_2]=i\theta \mathbb{I},
\end{equation}
where $\mathbb{I}$ is the $N \times N$ identity matrix. This is a non-commutativity constraint on the solutions of Eq.~\ref{SusskindMatrixModel} and is valid only for $N\to\infty$, as seen by taking a trace on both sides. The filling fraction $\nu$ is given by,
\begin{equation}
    \nu=\frac{2\pi\rho_0}{B}=\frac{1}{\theta B}
\end{equation}
In terms of the basis of matrix Hilbert space, $\ket{n},\;n=0,\cdots,N-1$, the system admits deformations as quasi-excitations. As an example, a quasi-hole excitation of charge $q>0$ at the origin appears in the non-commutativity constraint as
\begin{equation}
    [X_1,X_2]=i\theta(\mathbb{I}+q\ket{0}\bra{0}).
\end{equation}

To generalize to the case of finite $N$, Polychronakos introduced an additional boundary term, encoded in the complex vector $\Psi$~\cite{polychronakos2001quantum}. The Lagrangian for the counterpart finite $N$ MCS theory takes the following form,
\begin{align}
    L&=\frac{B}{2}\text{Tr}\left[\epsilon^{ab}(\Dot{X}_a-i[X_a,A_0])X_b+2\theta A_0\right]\nonumber\\
    &\qquad\qquad\qquad\qquad+\Psi^{\dagger}(i\partial_t-A_0)\Psi,
\end{align}
which possesses a $U(N)$ symmetry useful for later identification with the finite $N$ CSM \cite{calogero1971solution, PhysRevA.4.2019, Sutherland:1971ks}, corresponding to $X\to UXU^{\dagger},\;\Psi\to U\Psi$. 
An additional harmonic potential term can furthermore be added, of the form $-\omega X_a^2$, to deform the electron distribution of the droplet, concentrating electrons in the vicinity of the origin. 

Polychronakos has studied such finite $N$ MCS theories in planar~\cite{polychronakos2001quantum}, cylindrical~\cite{polychronakos2001quantumcyl}, and spherical systems~\cite{morariu_fractional_2005}, with analysis proceeding identically in each case save for minimal additional constraints. We therefore focus on the spherical case of greatest relevance to the present work, while briefly also touching upon some key results derived in the context of studying the cylindrical case.

In the spherical case, the model is first formulated in terms of stereographic coordinates corresponding to finite $N$ matrices $z$ and $z^{\dagger}$~\cite{morariu_fractional_2005}. While this formulation of the Lagrangian is important in discussion of quantization of the topological invariant $\text{Tr}[C]$ of the QSkHE~\cite{banerjee2024multiplicative, patil2024effective}, and is therefore shown in section~\ref{cquant}, it is not essential to quantization of MCS theory by its relation to Calogero integrable models. We therefore instead focus on the spherical MCS droplet formulated in terms of matrices $w$ and $w^{\dagger}$ parameterizing chord length from the north pole~\cite{morariu_fractional_2005}. This yields effectively the planar MCS Lagrangian,  
\begin{align}\label{lchord}
    \mathcal{L}(w,w^{\dagger})&=i\frac{B}{2}\text{Tr}\left[w^{\dagger}\Dot{w}-iA_0([w,w^{\dagger}]-2\theta)\right]\nonumber\\
    &-\frac{1}{2}w^2\text{Tr}[w^{\dagger}w]+\Psi^{\dagger}(i\Dot{\Psi}+A_0\Psi),
\end{align}
\textit{with the additional constraint due to spherical geometry}, that $w^{\dagger}w\leq4R^2$, where $R$ is the radius of the sphere. The eigenvalues of the matrix $w^{\dagger} w$ are a set of classical ``pseudoenergies'' $\{m_i\}$~\cite{morariu_fractional_2005}, which sum to the total energy. 

\begin{figure}[tbp]
    \centering
  \includegraphics[width=\columnwidth]{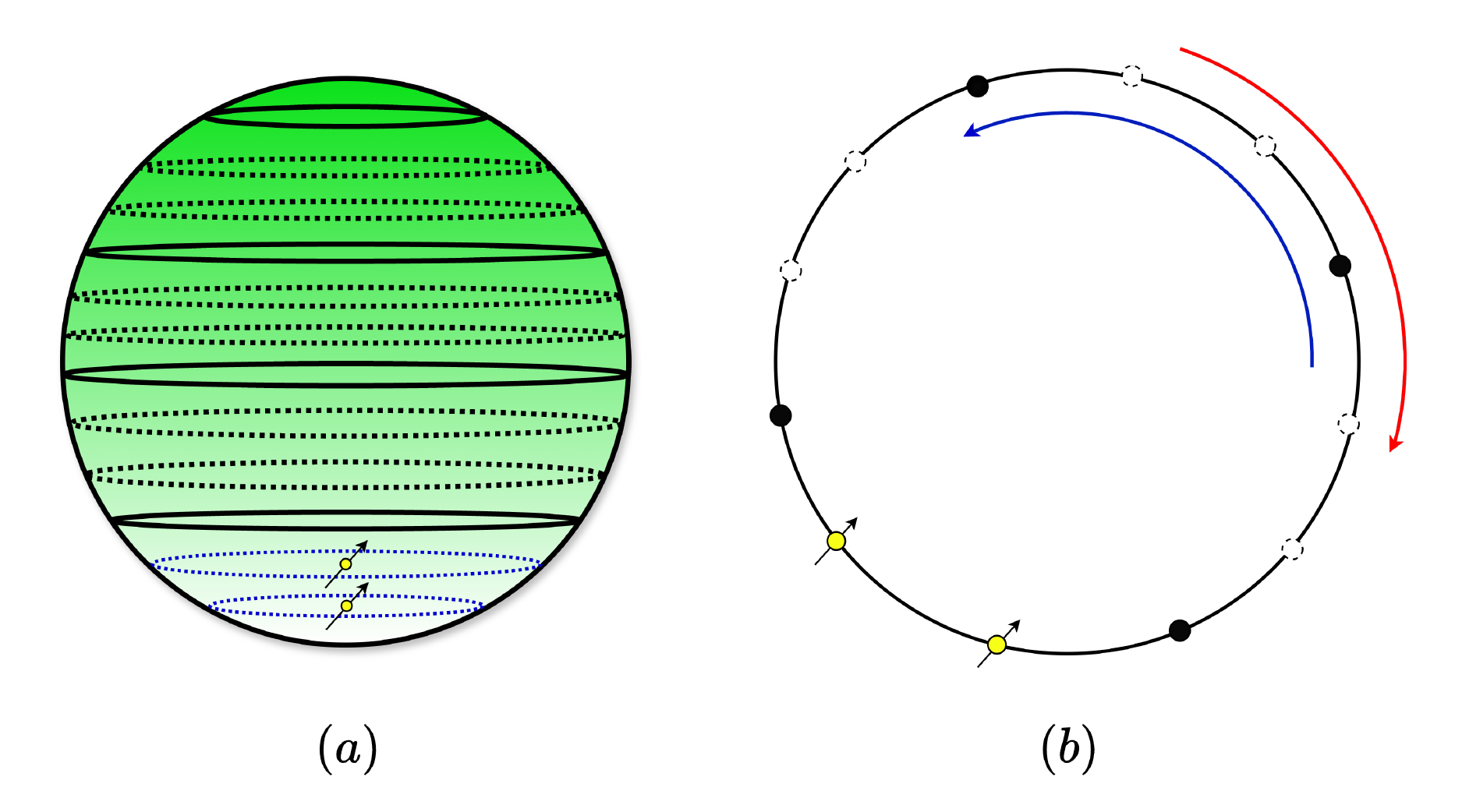}
  \caption{(a) Matrix Chern-Simons (MCS) theory for $N=4$ spinless electrons on a two-sphere, corresponding to matrix dimension of gauge fields of $N$, at filling fraction $1/(k+1)=1/3$. The Chern-Simons (CS) level of the infinite $N$ counterpart unbounded MCS theory of Susskind is $k=2$. States are filled by the electrons starting from the north pole, with four black circles representing the single-particle orbitals of a lowest Landau level (LLL) each occupied by an electron. Circles denoted by dashed black lines are unfilled orbitals, with two unfilled orbitals between each filled orbital as required by filling fraction $1/3$. Two circles in dashed blue lines depict a pair of bare flux quanta (each flux quantum is represented by small yellow circle with an arrow) near the south pole, corresponding to two unfilled levels consistent with $N=4$ and filling fraction $1/3$. (b) the MCS theory for $N=4$ spinless electrons at filling fraction $1/3$ is equivalently formulated in terms of a gauge theory for a Calogero-Sutherland model (CSM), a one-dimensional harmonic oscillator with periodic boundary conditions (PBCs) and an interaction term with coupling strength determined by the filling fraction $\nu$ of the MCS theory. The filled levels of the MCS theory are corresponding occupied sites of the CSM, shown as black filled circles. The unfilled levels of the quantum Hall (QH) droplet are represented by small, dashed-line circles. The two unfilled sites are represented by two yellow circles, each with an arrow.}
  \label{MCSandCSM_partialfill}
\end{figure}

To quantize the filling fraction $\nu$ of the finite $N$ MCS droplet for the purposes of the present work, it is important to note that this Lagrangian is also that of the $1+1$ D Calogero-Sutherland model (CSM)~\cite{morariu_fractional_2005}.  This is an integrable model for $N$ electrons on a line with periodic boundary conditions (PBCs). It is a useful formulation of the theory for the present work for being amenable to formulation of space-like Wilson lines and corresponding loops~\cite{polychronakos2001quantumcyl}. This equivalence between the MCS theory on a fuzzy sphere and the CSM for $N$ electrons is illustrated for a case of partial filling of the fuzzy sphere relevant to the present work, of $N=4$ and filling fraction $1/(k+1)$ with $k=2$ in Fig.~\ref{MCSandCSM_partialfill}.  

For the present work, it is important to discuss the physical significance of the CSM, which is more clearly illustrated for some readers by the related Sutherland model Hamiltonian. The Hamiltonian takes the form,
\begin{equation}\label{suthHam}
    H=\sum_{n=1}^N\frac{\omega}{2B}p_n^2+\sum_{n\neq m}\frac{\nu^{-2}}{4\sin^2\frac{(\phi_n-\phi_m)}{2}}
\end{equation}
with $(\phi_n,p_n)$ being the co-ordinates and momenta of particles on the circle and $\omega /B$ and $\nu$ being coupling constants. That is, the Hamiltonian is that of a quantum harmonic oscillator, with an additional interaction term governed by the filling fraction $\nu$ of the MCS theory.

The Gauss law, in terms of $X=\left(w + w^{\dagger} \right)/2$ and $P = B \left( w-w^{\dagger}\right)/2i$, takes the form~\cite{morariu_fractional_2005},
\begin{align}
\left[X,P\right] + i \Psi \Psi^{\dagger} = iB\theta,
\end{align}
with quantization of $B \theta$ to $k+1$ (this quantization will be discussed shortly), enforcing a constraint on the eigenvalues of the operator $w^{\dagger}w$, $\{m_i\}$. Eigenvalues of $X$ have the physical interpretation of particle coordinates, and $P$ encodes the conjugate momenta of the coordinates, in the basis in which $X$ is diagonal~\cite{morariu_fractional_2005}.

The $\{m_i\}$ are represented as lattice points with the constraint,
\begin{equation}
    m_{i+1}-m_i\leq k+1,
\end{equation}
The highest eigenvalue $m$, which is the monopole charge, corresponds to a state at the south pole defined by the spherical constraint as,
\begin{equation}
    4\pi BR^2=2\pi m.
\end{equation}

The filling fraction $\nu = 1/(k+1)$ is defined as the ratio of particle number $N$ to the number of possible states on the fuzzy sphere, an approximation of a sphere in terms of $N \times N$ matrices~\cite{Madore:1991bw}. The total number of possible states is defined in terms of the monopole charge $m$ as $L=m+1$ in the lowest Landau level (LLL),
\begin{equation}
    \nu=\frac{N}{L}
\end{equation}
where the saturation value $(N-1)(k+1)=m$ corresponds to the fully-filled sphere. More generally, $(N-1)(k+1)\leq m$. 

Quantization of $B\theta$ to $k+1$, and equivalently $\nu = 1/(k+1)$, proceeds from the Gauss law~\cite{POLYCHRONAKOS199129}. The Gauss law decomposes into what Polychronakos~\cite{polychronakos2001quantum} identifies as a ``traceless'' part,
\begin{align}
\left(G^a_{X} + G^a_{\psi}\right) |phys \rangle = 0,
\end{align}
with $|phys \rangle$ a quantum state, where $G^a_{X}$ and $G^a_{\psi}$ are of the form $G^a = a^{\dagger}_{\alpha} R^a_{\alpha \beta} a_{\beta}$, with $R^a_{\alpha \beta}$ the matrix elements of the generators of $SU(N)$. These generators are in a representation of dimension $d_R$, with $a_{\alpha}$, $a^{\dagger}_{\alpha}$ a set of $d_R$ mutually-commuting oscillators. $G^a_{\Psi}$ and $G^a_{X}$ are constructions for $R^a$ in the fundamental or adjoint representation, respectively. There is also what Polychronakos identifies as the ``trace'' part of the Gauss law,
\begin{align}
\left(\Psi^{\dagger}_n \Psi^{}_n - NB\theta \right)|phys \rangle = 0.
\end{align}
$\Psi^{\dagger}_n \Psi^{}_n$ is the total number operator for the oscillators and is therefore integer. $NB\theta$ must therefore be integer. The ``traceless'' part of the Gauss law, however, demands that the physical states be in a singlet representation of $G^a$. $G^a_X$ realizes representations of the adjoint, and so contains only irreducible representations (irreps) with a total number of boxes in their Young tableau, which are integer multiples of $N$. As the representations of $G_{\Psi}$ and $G_X$ must be conjugate to one another so their product contains the singlet, the irreps of $G_{\Psi}$ must also have a number of boxes in their Young tableau, which is a multiple of $N$.

Following earlier discussion by Polychronakos~\cite{POLYCHRONAKOS199129}, $G^a_{\Psi}$ is the Schwinger realization of the $SU(N)$ algebra in terms of oscillators, and therefore in the symmetric representation of $SU(N)$, labeled by the number operator $\Psi^{\dagger}_n \Psi^{}_n$. These irreps have quadratic Casimir $C_2 = \ell (\ell+1)$. Incorporating constraints on the action due to the Chern-Simons term then chooses the irreps with $\ell=k$. There is, as Polychronakos states, a resultant shift in the effective Chern-Simons level for finite $N$ from $k$ to $k+1$ (and corresponding change in filling fraction $\nu$), effectively due to the boundary term. We note that quantization also alters the interaction coupling of the CSM, with the replacement~\cite{POLYCHRONAKOS199129}
\begin{align}\label{quantSuth}
\nu^{-2} \rightarrow k(k+1).
\end{align}

We briefly discuss some of the deep connections of CSMs to other foundational topics of modern theoretical physics, to provide context to the reader for the significance of relating the QSkHE to the CSM later in the present work. We have already alluded to the connection between the CSM and the FQHE in stating the mapping by Polychronakos of the MCS Lagrangian to that of the CSM. There is extensive discussion in the literature further relating these two systems, however. The ground state of the CSM Hamiltonian is given by a one-dimensional Jastrow wave function, and the excitations can be obtained by multiplying Jack polynomials to the ground state \cite{ha_fractional_1995}. For odd filling fractions, such that $\nu=1/(2k+1),k\in I^+$, where $I^+$ is the set of positive integers, the ground and excited states of the FQHE can be projected to the edge, yielding the eigenstates of the CSM. Indeed, separate works detail the connection between the CSM and the fractional quantum Hall effect (FQHE), where the CSM appears as a microscopic model of the edge of a QH droplet at certain filling fractions ~\cite{kawakami_novel_1993,Yue_1998,yu_microscopic_1999}. The CSM is therefore also closely-related to work on conformal field theories (CFTs) as the boundary of a $2+1$ D topological phase can be described by a $1+1$ D CFT \cite{PhysRevB.43.11025, Tong:2016kpv, langmann_elliptic_2025, sakamoto_correspondence_2005, berntson_conformal_2025}. 

The CSM also captures the fractional \textit{exclusion} statistics of QH excitations, thus being a model realization of Haldane's generalized fractional statistics in any number of dimensions \cite{haldane_fractional_1991}. Past works~\cite{haldane_fractional_1991,ha_fractional_1995} have explored the relationship between fractional \textit{exchange} statistics - which is relevant for braiding of anyonic excitations in the $2+1$ D FQHE~\cite{Tong:2016kpv} - and fractional exclusion statistics, which are more relevant for $1+1$ D models such as the CSM \cite{ha_fractional_1995}. In a single spatial dimension, the exchange of particles must necessarily come with scattering events \cite{ha_fractional_1995}. However, exclusion statistics concerns itself with the Hilbert space structure of single particle excitations, and is specifically related to the change in the number of states $\Delta D$ for a given change in the number of excitations $\Delta N$, yielding the \textit{statistical parameter} $g=-\Delta D/\Delta N$. For fermions and bosons $g=1,0$ respectively, and for single particle excitations of the CSM, $g=1/(k+1)$~\cite{ha_fractional_1995} ($\lambda$ in ~\cite{ha_fractional_1995} is $k+1$ in our version of the model, Eqs.~\ref{suthHam},~\ref{quantSuth} after quantization). Notions of exchange and exclusion statistics can be studied via the CSM~\cite{ha_fractional_1995}.

 The CSM is also closely-related to a number of canonical Hamiltonians for spin chains. Following Haldane's generalization of fractionalized statistics to any number of dimensions~\cite{haldane_fractional_1991}, Bernard et. al. laid down the connection between the $XXX$ chain and a spinful CSM model~\cite{bernard_yang-baxter_1993}. The latter can be obtained from the $XXX$ chain by replacing lattice sites with dynamical particles, or particles with conjugate pairs of variables $(\phi,p)$, such as for real-space position and momentum~\cite{bernard_yang-baxter_1993}. The spinful CSM is also closely-related to the well-known Haldane-Shastry model~\cite{PhysRevLett.60.635,PhysRevLett.60.639,pasquier2005lecture,ha_fractional_1995}, which is the so-called 'freezing limit' of the spinful CSM \cite{lamers_fermionic_2024}. The $XXX$ chain, the Haldane-Shastry model and the spinful/spinless CSM all exhibit the generalized Pauli exclusion principle first noted by Haldane~\cite{haldane_fractional_1991}. Considering that these systems are all examples of integrable systems, relevant concepts such as the Yang-Baxter equation and the Bethe ansatz have also been applied and studied in the context of the CSM ~\cite{bernard_yang-baxter_1993, ferrando_bethe_2025, enciso_spin_2009}. 
 
 In addition to the strong association of the CSM with the FQHE and spin chains, the CSM appears prominently in discussion, for example, of circular ensembles in random matrix theory ~\cite{dyson_statistical_1962}. The ground state wavefunctions of the CSM at $\lambda = 1/2,1,2$ ($k+1$ in Eqs.~\ref{suthHam},~\ref{quantSuth} is replaced by $\lambda$), respectively, can be mapped to the eigenvalue distributions of the orthogonal, unitary and symplectic random matrices respectively \cite{ha_fractional_1995,taniguchi_random_1995}.

We close this brief review by highlighting that results on finite $N$ MCS theory hold for general $N$. This includes $N$ small as for cases of DOFs associated with spin operator matrix representations of dimension $N$ normally interpreted as isospin, or internal, DOFs, although MCS theory has not previously been considered in such scenarios. In subsequent sections, we will highlight some important consequences of MCS in this regime, and also present some essential generalizations of MCS theory that strengthen the case for applying MCS theory at small $N$, both to more accurately model an individual quantum spin $S$ of multiplicity $N=2S+1$ and to construct microscopic theories of the QSkHE.

\subsection{Brief review of the quantum skyrmion Hall effect, focusing on the multiplicative Chern insulator}

Here, we will briefly summarize the QSkHE and the response of the MCI that most directly motivates the present work. At the level of effective field theories presented in Patil~\emph{et al.}~\cite{patil2024effective} consistent with many related and earlier works~\cite{qskhe, cook2022multiplicative, cook2023topological, PhysRevB.108.045144, ay2024signatures, winter2025observable, banerjee2024multiplicative}, the QSkHE is the collection of phenomena associated with Hamiltonians with spin DOFs, with corresponding spin operators of matrix dimension $N$, potentially encoding a finite number of spatial dimensions both for $N$ large~\cite{aschieri2007, aschieri2004dimensional, Aschieri:2004vh, Chatzistavrakidis:2010tq} but also for $N$ small, in the regime where the spin DOF has previously been treated as an isospin, or internal DOF. 

A variety of signatures of the QSkHE are reported in lattice tight-binding model Hamiltonians quadratic in second-quantized creation and annihilation operators~\cite{ay2024signatures,banerjee2024multiplicative}, with some results even suggesting first experimental observation of the QSkHE in past experiments on HgTe quantum wells~\cite{ma2015unexpected, bernevig2006quantum}. These signatures in lattice tight-binding models are associated with three sets of topologically non-trivial phases of matter, which are lattice counterparts of the QSkHE~\cite{cook2022multiplicative,cook2023topological,PhysRevB.108.045144}, and demonstrate that the spin DOF associated with matrix dimension $N$ small yielding signatures of the QSkHE could have various physical meanings, including the physical spin, orbital angular momentum, or layer (pseudo)spin DOF.  

A minimal action for the QSkHE at the level of effective field theories illustrates the essential generalization. The action is formulated in terms of $D$ Cartesian spatial coordinates and $\delta$ fuzzy dimensions, with $D$ and $\delta$ each finite. As a result, the action is distinguished by terms with gauge fields over extra fuzzy dimensions defined in terms of spin DOF(s) encoding the $\delta$ dimensions. The corresponding spin operators are of matrix dimension $N$ small such that it would typically be interpreted as an isospin~\cite{aschieri2007, aschieri2004dimensional, Aschieri:2004vh, Chatzistavrakidis:2010tq, patil2024effective}. A characteristic additional term in the action, for a system with two Cartesian space coordinates ($D=2$), two dimensions encoded in spin $\delta=2$ and one time coordinate is then,
\begin{align}
    S_{CS}=&C_2\int d^3x \text{kTr}\left[\text{Tr}(G)\;CS_3\wedge \Hat{F}\right],\\
   CS_3=&\left[AdA+\frac{2}{3}AAA\right],\\
   \Hat{F}= & \left[ X_a, A_b \right]   - \left[ X_b, A_a \right] + \left[A_a, A_b \right] - C^c_{ab} A_c,
\end{align}

\begin{figure}[tbp]
    \centering
    \includegraphics[width=\columnwidth]{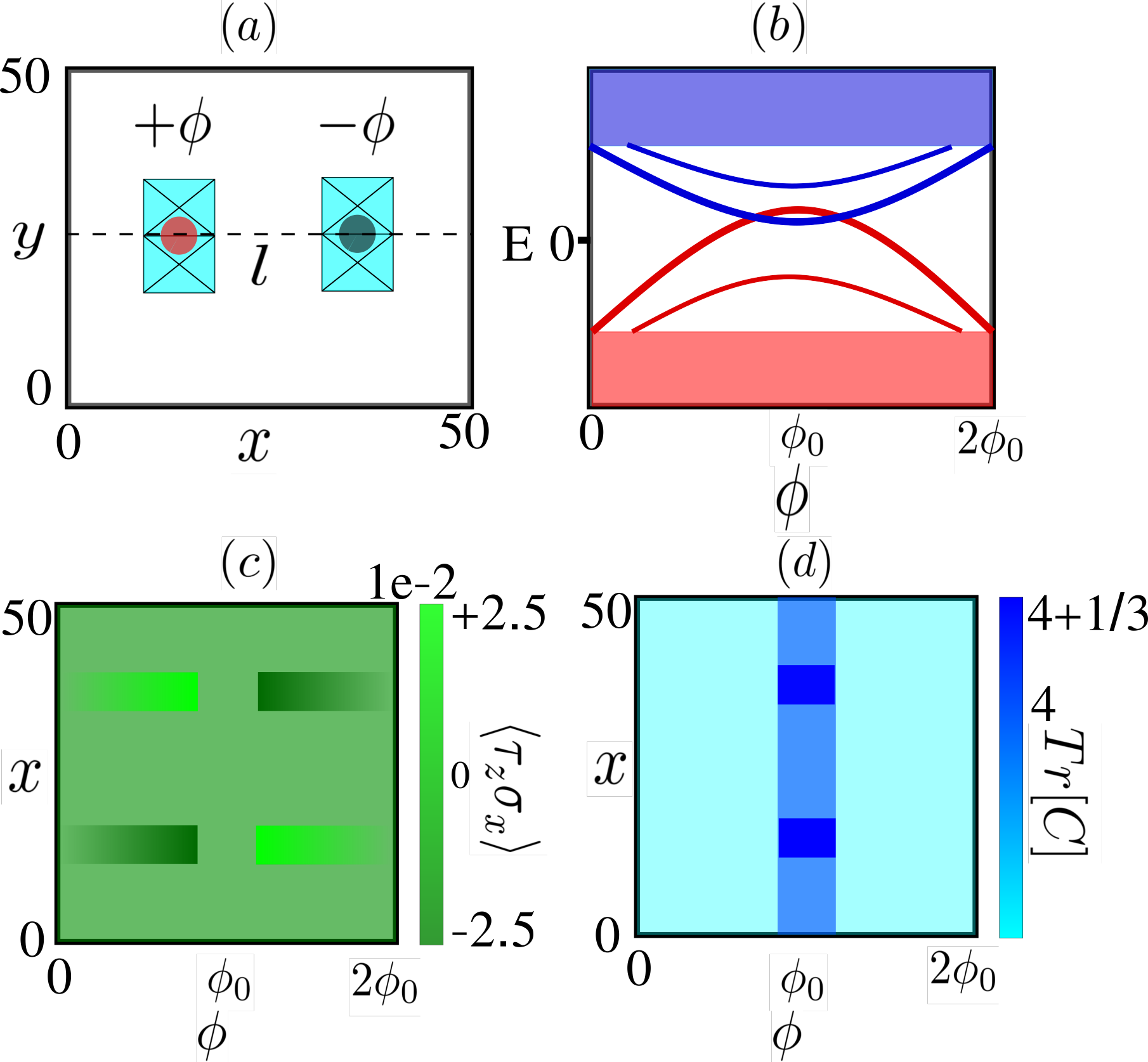}
    \caption{Schematic summary of results for the time-reversal invariant flux insertion response of the MCI as discussed in~\cite{banerjee2024multiplicative}. (a) Flux insertion schematic, with red and grey circles denoting flux tubes carrying fluxes of equal magnitude and opposite in sign, with a line $l$ of lattice sites intersecting the two flux tubes; (b) Sketch of the energy spectrum, with states entering the gap from the set of bulk conduction bands and valence band states colored blue and red, respectively. Note that the level-crossing of the red and blue states lies above zero energy, although the system is particle-hole symmetric for zero flux; (c) Sketch of the real-space position-resolved values of $\left<\tau_z\sigma_x\right>$ for a cut in real-space along the line $l$ (at $y=y_0=25$) as a function of inserted flux $\phi$; (d) Sketch of the real-space position resolved values of $\text{Tr}[C]$ along the line $l$ as a function of inserted flux, showing a value of $4+1/3$ for the interval in flux $\phi$ around $\phi=\phi_0$ bounded by level-crossings shown in (b).}
    \label{fig:Fig1MCI}
\end{figure}

where $C_2$ is the second Chern number, $CS_3$ is the standard 2+1 D non-Abelian CS Lagrange density, and $\hat{F}$ is the field strength defined over a fuzzy coset space isomorphic to the fuzzy two sphere $S^2_F$~\cite{qskhe, aschieri2007, aschieri2004dimensional, Aschieri:2004vh, Chatzistavrakidis:2010tq, patil2024effective}. $\text{Tr}(G)$ is the trace over the gauge group $G$, and $\text{kTr}$ is the normalized trace over fuzzy coset space coordinates. The field strength over the fuzzy space coordinates is expressed in terms of $N \times N$ matrix Lie algebra generators $\{X_a\}$, gauge field components $\{A_a\}$, and structure factor $C_{abc}$ (a generalization of a structure constant~\cite{aschieri2007, aschieri2004dimensional, Aschieri:2004vh, Chatzistavrakidis:2010tq}) and notably includes an additional term associated with the formulation of a Lie derivative on the fuzzy sphere~\cite{Madore:1991bw} compared to the commutative non-Abelian formulation of the field strength.

In cases of such additional terms in an action, phenomenology of the QSkHE is similar to that of the 4+1 D QHE and Chern insulator~\cite{Hu:2001kf, Bernevig:2002eq, Qi:2008ew, patil2024effective}, with replacement of Landau levels (LLs) by \textit{severely-fuzzified LLs}, or LL\textsubscript{F}s~\cite{patil2024effective}. These LL\textsubscript{F}s are potentially almost point-like LLs defining the extra fuzzy dimensions, which, in this case, can still host intrinsically $2+1$ D topological states identified with quantum skyrmions. Such quantum skyrmions are non-commutative counterparts of e.g., the skyrmions which are low-lying charged excitations of the quantum Hall ferromagnet (QHFM) in a commutative treatment~\cite{PhysRevB.47.16419}. Propagating LL\textsubscript{F}s at the boundary are also understood as severely-fuzzified three-dimensional Weyl nodes, or WN\textsubscript{F}s~\cite{Qi:2008ew}.

Importantly, such quantum skyrmions have been characterized in numerics using a proposed topological invariant, $\text{Tr}[C]$, which derives from assumption of flat connection over the matrix (fuzzy) space as 
\begin{align}
    \hat{F}_{ab}=0=\left[A_a, A_b \right] - C_{abc}A_c.
\end{align}
The \textit{structure factor} $C_{abc}$~\cite{patil2024effective} is then expressed in terms of the gauge field components $\{A_i\}$. $\text{Tr}[C_{abc}]$ serves as a topological invariant, which may be computed for individual unit cells of the lattice tight-binding models, with each unit cell formulated in terms of an $N \times N$ matrix. In numerics, more specifically, $C_{abc}$ is computed as the quantity $C^{occ}_{abc}$~\cite{banerjee2024multiplicative} from spin operators $\{S_i\}$ projected to an occupied subspace defined by a density matrix $\rho$, or projected spin operators $S^{occ}_i = \rho S_i$, as
\begin{align}
    \left[S^{occ}_a, S^{occ}_b \right] = C^{occ}_{abc}S^{occ}_c.
\end{align}
This invariant has been used to characterize the Bernevig-Hughes-Zhang model~\cite{bernevig2006quantum, ay2024signatures} and the Hamiltonian for the multiplicative Chern insulator (MCI)~\cite{banerjee2024multiplicative}. For the present work, it is important to summarize how $\text{Tr}[C]$ is used to characterize the topological response of the MCI specifically.

The MCI is one of the topologically non-trivial phases of matter related to the QSkHE, which are known as multiplicative topological phases ~\cite{cook2022multiplicative, banerjee2024multiplicative}. These are phases of matter first realized in lattice tight-binding Hamiltonians, characterized by the Hamiltonians possessing a symmetry-protected tensor product structure~\cite{cook2022multiplicative}. Examples studied thus far are for Bloch Hamiltonians with $4\times 4$ matrix representation ($N=4$) possessing a symmetry-protected Kronecker product structure, such that they are expressed as a Kronecker product of two 'parent' Bloch Hamiltonians, $\mc{H}_{p1}(\boldsymbol{k}_1) = \mathbf{d}_1 (\boldsymbol{k}_1)\cdot \boldsymbol{\tau}$ and $\mc{H}_{p2}(\boldsymbol{k}_2) = \mathbf{d}_2 (\boldsymbol{k}_2) \cdot \boldsymbol{\sigma}$, which each have $2 \times 2$ matrix representation, or $N=2$, with $\boldsymbol{\tau}$ and $\boldsymbol{\sigma}$ each being a three-vector of Pauli matrices, and $\mathbf{d}_1$ and $\mathbf{d}_2$ being two three-vectors of momentum-dependent functions. The $4 \times 4$ child Bloch Hamiltonian is therefore expressed as
\begin{align}
    \mathcal{H}_c &= \sum_{\boldsymbol{k}_1, \boldsymbol{k}_2} \Psi^{\dagger}_{\boldsymbol{k}_1, \boldsymbol{k}_2}\mathcal{H}^c(\boldsymbol{k}_1, \boldsymbol{k}_2) \Psi^{}_{\boldsymbol{k}_1, \boldsymbol{k}_2} \nonumber \\
    &= \sum_{\boldsymbol{k}_1, \boldsymbol{k}_2} \Psi^{\dagger}_{\boldsymbol{k}_1, \boldsymbol{k}_2}[\mc{H}_{p1}(\boldsymbol{k}_1) \otimes \mc{H}_{p2}(\boldsymbol{k}_2)] \Psi^{}_{\boldsymbol{k}_1, \boldsymbol{k}_2},
\end{align}
where the basis vector is $\Psi_{\boldsymbol{k}_1, \boldsymbol{k}_2} = \left( c^{}_{\boldsymbol{k}_1, \boldsymbol{k}_2, \uparrow}, c^{}_{\boldsymbol{k}_1, \boldsymbol{k}_2, \downarrow}, c^{\dagger}_{-\boldsymbol{k}_1, -\boldsymbol{k}_2, \downarrow}, c^{\dagger}_{-\boldsymbol{k}_1, -\boldsymbol{k}_2, \uparrow}\right)^{\top}$, where $c^{\dagger}_{\boldsymbol{k}_1, \boldsymbol{k}_2, \sigma}$ creates a fermion labeled by two momenta $\boldsymbol{k}_1$ and $\boldsymbol{k}_2$ and spin $\sigma \in \{\uparrow, \downarrow \}$. The two parent Bloch Hamiltonians and the child Bloch Hamiltonian take the forms 
\begin{align}
    \mc{H}_{p1}(\boldsymbol{k}) &= [m-2t\left( \cos(k_x) + \cos(k_y) \right)] \tau_z \nonumber \\
    &+ \Delta  \sin{k_x} \tau_x + \Delta \sin{k_y} \tau_y \\
    \mc{H}_{p2}(\boldsymbol{k}) &= \mathcal{T} \mathcal{H}_{p1}(\bk)  \mathcal{T}^{-1} \\
    \mathcal{H}^c(\boldsymbol{k}) &= \mc{H}_{p1}(\boldsymbol{k}) \otimes \mc{H}_{p2}(\boldsymbol{k}),
\end{align}
 where $\mathcal{T}=i \tau_y \mathcal{K}$ is the spinful time-reversal operator, with $\mathcal{K}$ being complex conjugation and $\boldsymbol{\tau}$ changed to $\boldsymbol{\sigma}$ after time-reversal is performed.

 The child Hamiltonian $\mathcal{H}^c(\boldsymbol{k})$ constructed as a Kronecker product of the two parent Hamiltonians inherits various properties from its parent Hamiltonians: the bulk spectrum, as a simple example, consists of eigenvalues which are products of eigenvalues of the two parents. Phenomenology of multiplicative topological phases becomes far richer, however, with controlled breaking of the tensor product structure, such as by open boundary conditions (OBCs). OBCs yield bulk-boundary correspondence well-understood in terms of the bulk-boundary correspondences of the parent topological phases, with entanglement however yielding additional features beyond those of either individual parent.

For the present work, we specifically review the topological response of the MCI ~\cite{banerjee2024multiplicative}. Banerjee~\emph{et al.} ~\cite{banerjee2024multiplicative} presents characterization of the response of the MCI to time-reversal invariant (TRI) insertion of magnetic flux through two spatially well-separated plaquettes of the MCI lattice. A schematic of this flux insertion process for a system size of $50$ by $50$, with periodic boundary conditions (PBCs) in each of the $\hat{x}$ and $\hat{y}$ directions in real-space is shown in Fig.~\ref{fig:Fig1MCI}(a). Red and grey dots highlight the locations at which flux $+\phi$ and $-\phi$ are inserted, each depicted with two plaquettes of the lattice enlarged for illustrative purposes. A second schematic of results in Fig.~\ref{fig:Fig1MCI}(b) illustrates dependence of spectrum on flux $\phi$ inserted in the manner of Fig.~\ref{fig:Fig1MCI}(a). The MCI with PBCs subjected to such a flux insertion is $2\phi_0$ periodic, where $\phi_0$ is the flux quantum, and the spectrum depicts reported level crossings (between blue and red states) in the vicinity of $\phi = \phi_0$. 

Correspondingly, the evolution of the ground state under such flux insertion (with Fermi level intersecting the two level crossings near $\phi=\phi_0$) was characterized by computing the expectation value of a (pseudo)spin operator with matrix representation $\tau_z \sigma_x$ vs. flux $\phi$ and position $x$ in real-space for the real-space cut at $y=y_0=25$. The evolution of this spin texture vs. flux $\phi$ is illustrated in Fig.~\ref{fig:Fig1MCI}(c). The texture is $2\phi_0$-periodic. Additionally, a topological invariant associated with the parent isospin DOFs was computed ~\cite{banerjee2024multiplicative, patil2024effective}. In notation of the present work, the invariant is computed for each unit cell in real-space at location $\left(x,y\right)$ and given by:

\begin{align}
    \text{Tr} \left[C^{occ}(x,y)\right]=\text{Tr} \left[\left[ S^{occ}_{x}(x,y) , S^{occ}_{y}(x,y) \right] \left(S^{occ}_{z}(x,y)\right)^{-1}\right]
\end{align}
Where $S^{occ}_{i}(x,y) = \rho(x,y) \tau_i \sigma_0$ and $\rho(x,y)$ is the $4 \times 4$  block of the projector-onto-occupied states of the full system, $\rho$, for the unit cell at position $(x,y)$, with $\rho$ computed from the one-point correlator according to Peschel's method~\cite{IngoPeschel_2003}. Fig.~\ref{fig:Fig1MCI}(d) shows  $\text{Tr} \left[C^{occ}(x,y) \right]$, abbreviated to $\text{Tr} \left[\boldsymbol{C} \right]$ in Fig.~\ref{fig:Fig1MCI}, vs. magnetic flux $\phi$ and layer index in the $\hat{x}$-direction, $x$, for a cut in real-space along the line $l$ shown in Fig.~\ref{fig:Fig1MCI}(a) at $y=y_0=25$. At sites of flux insertion for $\phi$ near $\phi_0$, $\text{Tr} \left[\boldsymbol{C} \right]$ deviates from $4$, the value otherwise observed for $\phi$ in the interval between the level crossings, by very close to $1/3$ in numerics: at the level crossing for $\phi<\phi_0$, quantization is effectively exact, and deviates slightly with increasing $\phi$ as states carrying the $1/3$ contribution delocalize and merge with the bulk as characterized by inverse participation ratio. The signatures of the MCI topological response most essential to the present work are therefore summarized as a $4 \pi$ Aharonov-Bohm (AB) effect, associated with Chern-Simons level $k=2$ and filling fraction $\nu = 1/2$, occurring in combination with a contribution of $1/3$ to $\text{Tr} \left[\boldsymbol{C} \right]$ at sites of flux insertion for flux $\phi$ close in value to $\phi_0$.

\section{Generalized quantization procedure in the context of matrix Chern-Simons theory}\label{cquant}
In this section, we present extensions of MCS theory for a single MCS/CSM droplet, or pair of MCS/CSM droplets, for cases where the number of electrons in each droplet is small. We specifically apply these results to cases of a single $N=4$ MCS/CSM droplet or a pair of $N=2$ MCS/CSM droplets. These results relate MCS theory to study of individual unit cells of the lattice tight-binding Hamiltonians realizing signatures of the QSkHE, which will later be generalized to describe arrays of multiple unit cells.

\begin{figure}[tbp]
    \centering
  \includegraphics[width=\columnwidth]{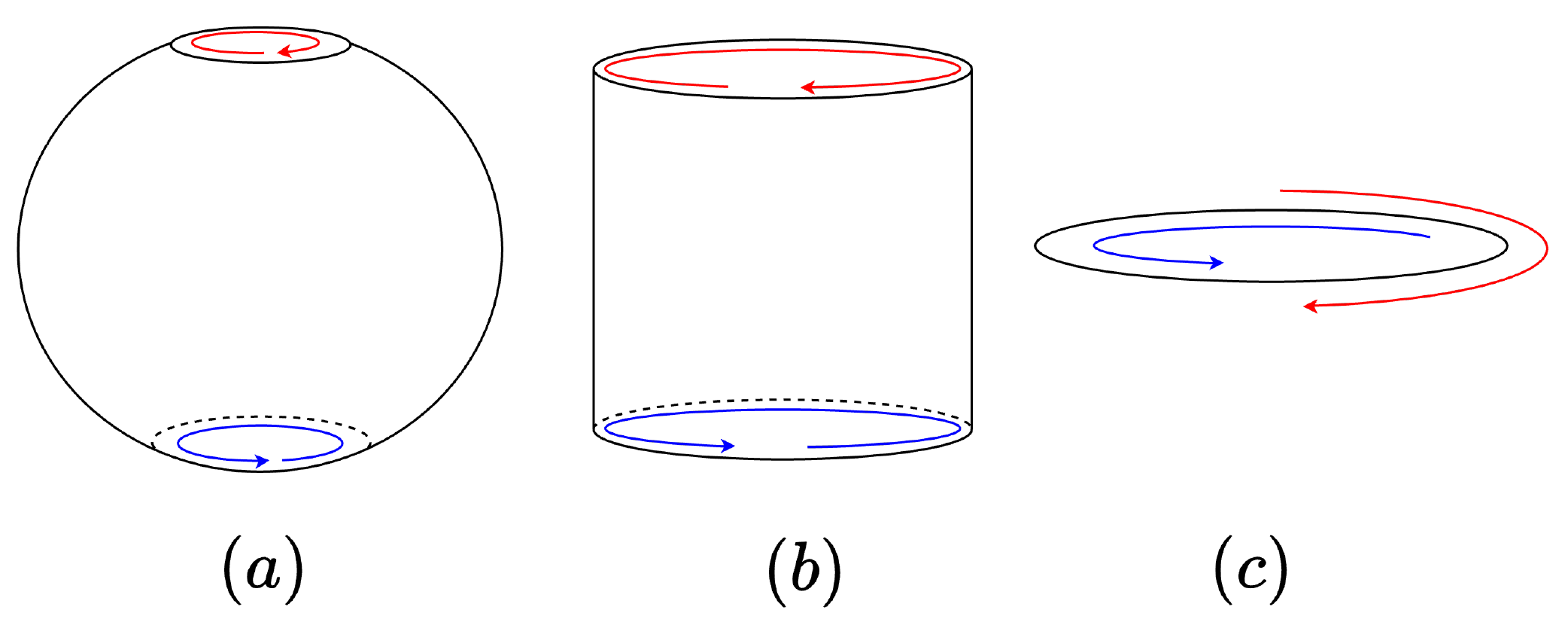}
  \caption{Schematics of the finite $N$ MCS theory defined on (a) a two-sphere and (b) a cylinder, with the counterpart CSM theory depicted schematically in (c). Red and blue arrows depict chiral edge modes of the FQH state effectively modeled by the MCS theory, with the theory on the sphere being equivalent to that on the cylinder up to an additional truncation of the Hilbert space. This truncation may be thought of as pinching the open edges of the cylinder to point-like counterparts near the north and south poles of the sphere. The counterpart CSM models the two chiral edge modes together in a 1D system, in terms of fractional exclusion statistics and scattering modeled by an interaction term with strength determined by the filling fraction of the counterpart MCS theory.}
  \label{spherecylindercalogero}
\end{figure}

We begin this section by generalizing the quantization procedure within MCS theory and relate this result to the topological invariant $\text{Tr}[C]$ of the QSkHE. This discussion collects together results for bounded MCS theory (with matrix dimension $N$ finite) on the sphere and cylinder and the counterpart CSM, which are closely-related. The equivalence is summarized in the schematic Fig.~\ref{spherecylindercalogero}.

We start from the spherical bounded matrix model in stereographic coordinates presented by Morariu and Polychronakos~\cite{morariu_fractional_2005}. In terms of $N\times N$ matrix gauge field components $z$ and $z^{\dagger}$, the Lagrangian takes the form,
\begin{align}\label{sphereL}
    \mathcal{L}(z, z^{\dagger}) &= i B \text{Tr}\left[2R^2\left(1+z z^{\dagger}\dot{z} z^{\dagger} \right)^{-1}\right. \\ \nonumber
    &\left.- i A_0\left(2 R^2 \left[z, z^{\dagger}\left(1+z z^{\dagger} \right)^{-1} \right] - \theta\right) \right]\\ \nonumber
    &+ \Psi^{\dagger}\left(i \dot{\Psi} + A_0 \Psi \right)
\end{align}
where $R$ is the radius of the fuzzy sphere  and $\Psi$ is the complex column $N$ vector defining the boundary for the finite $N$ MCS Lagrangian as discussed in section ~\ref{secrew}. We note here that Polychronakos introduces a quantum ordering convention, in which $z$ and $z^{\dagger}$ alternate in any expressions containing these objects.

The Gauss law for the Lagrangian Eq.~\ref{sphereL}  identified previously is
\begin{equation}\label{zEOM}
    2BR^2 \left[z, z^{\dagger} \left(1+z z^{\dagger} \right)^{-1} \right] + \Psi \Psi^{\dagger} = B \theta.
\end{equation}

For our purposes, it is important to include counterpart formulations of the Lagrangian Eq.~\ref{sphereL}, which are also more directly identified as the Lagrangian for the CSM. Introducing $w = 2Rz \left(1+z^{\dagger}z \right)^{-1/2}$, we may instead write the Lagrangian as in the review section ~\ref{secrew}  on MCS theory, as
\begin{align}
\mathcal{L}(w, w^{\dagger}) &= i \frac{B}{2} \text{Tr}\left[w^{\dagger} \dot{w} - iA_0 \left(\left[w, w^{\dagger} \right]-2\theta \right) \right]\nonumber \\ 
&+ \Psi^{\dagger}\left(i \dot{\Psi} + A_0 \Psi \right),
\end{align}
with the spherical constraint $w^{\dagger}w \leq 4 R^2$~\cite{morariu_fractional_2005}. The solution and quantization of this model is that of the planar and cylindrical models discussed in section ~\ref{secrew}, save for the additional truncation due to the spherical constraint. Here, eigenvalues of $X=\left(w + w^{\dagger} \right)/2$ are particle coordinates, and $P = B \left( w-w^{\dagger}\right)/2i$ encodes the conjugate momenta of the coordinates, in the basis in which $X$ is diagonal~\cite{morariu_fractional_2005}.

For our discussion, it is useful to highlight an additional reformulation of the Lagrangian by Polychronakos that is specifically also used for discussion of the unitary matrix model for quantum Hall states on the cylinder. For this purpose, we take $z=hU$~\cite{morariu_fractional_2005}, with 
\begin{align}
    U&=\left(z z^{\dagger} \right)^{-1/2}z \\
    h&=\left(z z^{\dagger} \right)^{1/2} \\
    H &= \left(h^2-1 \right)\left(h^2 +1 \right)^{-1} \\
    P &= BR^2 H,
\end{align}
and express the Lagrangian as
\begin{align}
\mathcal{L}\left(U, H\right) &= iBR^2 \text{Tr}\left[H \dot{U} U^{-1} - i A_0 \left(\left[U,H\right] - \theta/R^2 \right) \right]\nonumber \\ 
&\qquad+\Psi^{\dagger} \left(i \dot{\Psi} + A_0 \Psi \right).
\end{align}
In this formulation, the spherical constraint manifests as restriction of eigenvalues of $H$ to lie within the interval $\left[-1, 1 \right]$~\cite{morariu_fractional_2005}.

The Gauss law is equivalently written as
\begin{align}
   \left[U, P \right] + \Psi \Psi^{\dagger} = B \theta.
    \label{UPEOM}\end{align}

In this notation, Polychronakos identifies $\left[U,P \right]$ as the sum of the spatial Wilson line operators for right rotation $\mathcal{R}$ and left rotation $\mathcal{L}$, or $G_U = \cal{L} + \cal{R}$~\cite{polychronakos2001quantumcyl}. As such, $2BR^2 \left[ z, z^{\dagger}\left(1+z z^{\dagger}\right)^{-1}\right]$, $\left[X,P \right]$, $\left[U,P \right]$ are all expressions for two spatial Wilson lines. 

While $G_U$ is identified as traceless by Polychronakos and collaborators based on the quantization procedure they use following Susskind ~\cite{susskind2001quantum}, this quantization procedure does not account for the form of the Lie derivative on matrix spaces discussed most explicitly, perhaps, in the context of fuzzy sphere geometry~\cite{aschieri2006dynamical, Madore:1991bw}. The Gauss law for the MCS theory on the sphere is derived by constructing the quantum action through conversion of Poisson brackets, defined in Eq.~\ref{poisson}, to a commutator, a standard procedure. This procedure, however, is potentially inconsistent with differential geometry of fuzzy coset spaces. As stated in work by Madore and others on gauge theories of fuzzy spheres~\cite{balachandran2005lectures, Madore:1991bw, morariu_fractional_2005}, the two-form on a fuzzy sphere takes into account that a partial derivative generalizes to a Lie derivative, as
\begin{align}
F_{ab} &= [X_a, A_b]-[X_b,A_a] + [A_a, A_b] - C_{abc}A_c,
\label{fuzzycurv}
\end{align}
where $\{X_i\}$ and $\{A_i\}$ are matrix valued co-ordinates and gauge fields over the fuzzy sphere, respectively, and $C_{abc}$ are the structure constants.

In the context of partial filling of the fuzzy sphere~\cite{morariu_fractional_2005}, the matrix gauge fields define some deformation from fuzzy sphere geometry, which can be encoded by generalizing from the structure constants $C_{abc}$ to the structure factors  $C^{occ}_{abc}$ introduced in Patil~\emph{et al.}~\cite{patil2024effective}. The structure factors encode potential algebraic artifacts of projection to occupied subspaces, such as partial filling of the fuzzy sphere~\cite{morariu_fractional_2005}. 

A minimal statement of flat connection associated with the Gauss law corresponds to the two-form $F_{ab}$, the non-commutative curvature associated with the gauge fields, being zero, or $F_{ab}=0$. As well, there should be no variation in the gauge fields, so the commutators $[X_a, A_b]$ and $[X_b, A_a]$, which may be identified with $\partial_a A_b$ and $\partial_b A_a$ in a commutative, non-Abelian CS theory, are also zero, or $[X_a, A_b]=[X_b, A_a]=0$. Equivalently, we could rewrite $F_{ab}$ in terms of $\phi_a = A_a - X_a$, yielding $F_{ab}= [\phi_a, \phi_b] - C_{abc}\phi_c$~\cite{Madore:1991bw} to simplify discussion. The expression of the curvature $F_{ab}$ on a partially-filled fuzzy sphere or other fuzzy space for flat connection (vanishing curvature) is then 
\begin{align}
    F_{ab} = 0 = [A_a, A_b] - C^{occ}_{abc}A_c.
    \label{flatconn}
\end{align}
This relation Eq.~\ref{flatconn} can then be incorporated by substitution into the Gauss law for the MCS theory on the fuzzy sphere as formulated in Morariu and Polychronakos~\cite{morariu_fractional_2005}. Identifying the commutators in $G_U$ with $\left[A_a, A_b\right]$ in the statement of flat connection above, these expressions for two Wilson lines are no longer traceless in general, as they can be substituted for a term of the form $C^{occ}_{abc}A_c$ representing a third spatial Wilson line $A_c$ dictated by previously unidentified fusion rules encoded in $C^{occ}_{abc}$.

Importantly, this result Eq.~\ref{flatconn} is analogous to a minimal non-trivial scenario of commutative, non-Abelian CS theory, for insertion of three Wilson lines corresponding to $SU(N)_{k=2}$~\cite{Tong:2016kpv}. In this case, Wilson lines $a$, $b$, and $c$ satisfy non-trivial fusion rules according to $a \star b = \sum_c N^c_{ab}c$. Correspondingly, rather than considering Young tableaux for $R \times \bar{R}$ for Wilson lines $\mathcal{L}+\mathcal{R}$, we must generalize, as a result of the differential geometry of the partially-filled fuzzy sphere, to $R \star \bar{R}$, and determine the fusion rules for a generalization from non-Abelian anyons of a commutative, non-Abelian CS theory, to anyons of a bounded MCS theory.

\begin{figure}[tbp]
    \centering
  \includegraphics[width=\columnwidth]{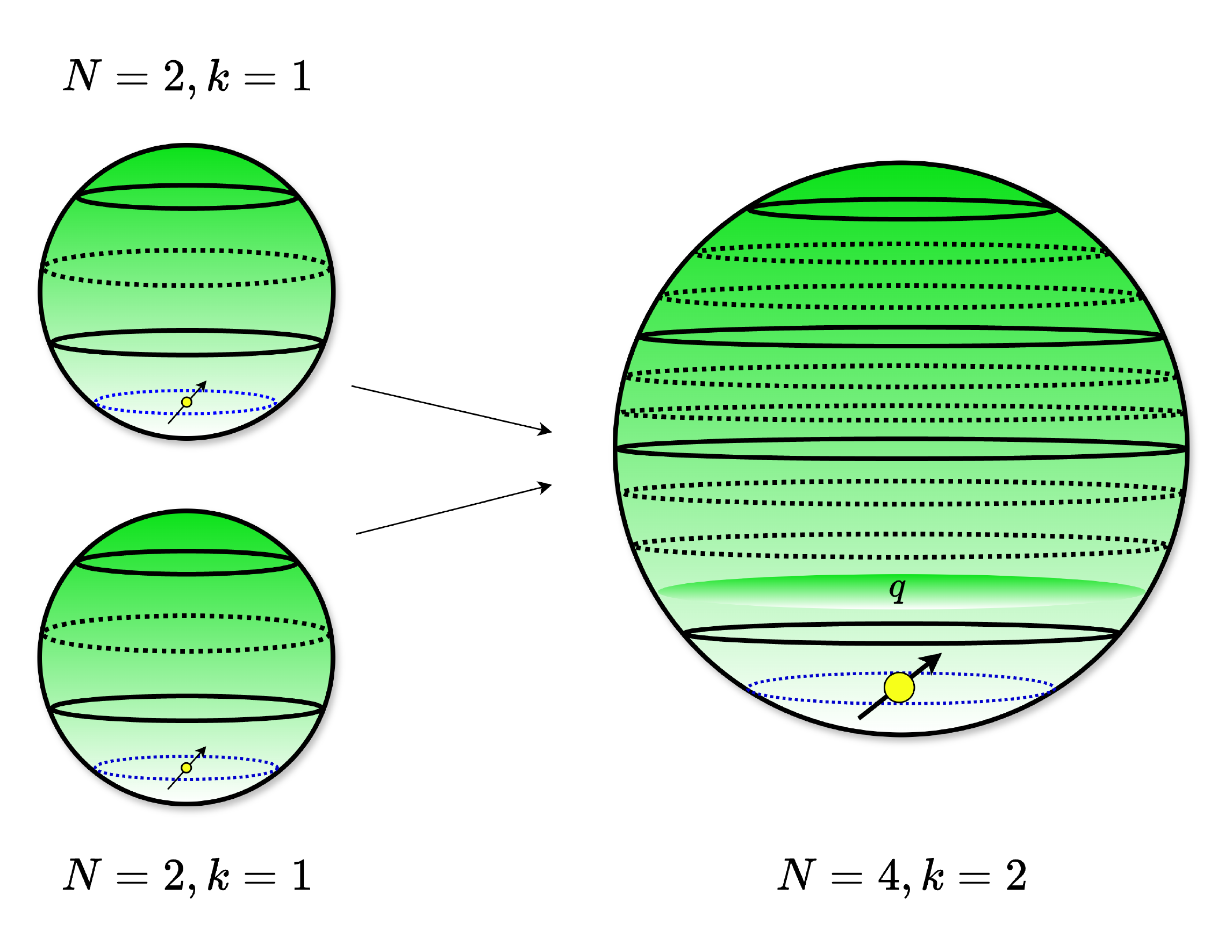}
  \caption{Schematic of merging process consistent with signatures of the topological response of the MCI: two $N=2$ MCS theories---one for each parent Hamiltonian contribution to the unit cell of the MCI---defined as partially-filled, deformed fuzzy spheres, each with a single bare level near the south pole. This corresponds to filling fraction $1/2$ for each $N=2$ droplet, consistent with the total Chern numbers of $\pm 1$ of the parent Hamiltonians modeled as unbounded MCS theories at level $k=1$. Two arrows point from these $N=2$ droplets to a single $N=4$ MCS droplet at filling fraction $1/(2+1)=1/3$, representing evolution of a single unit cell of the MCI during flux insertion. While this filling fraction for $N=4$ is consistent with two bare levels as shown in Fig.~\ref{MCSandCSM_partialfill} in the absence of quasi-particle/hole excitations, flux insertion through the MCS droplet on the deformed fuzzy sphere also yields a quantum skyrmion excitation, modeled here as the simplest example of a quasi-hole excitation with charge $1/3$, corresponding to the $1/3$ contribution to $\text{Tr}[C]$ observed in numerics for the MCI.}
  \label{fillfrac_merge}
\end{figure}

As the Gauss law demands a singlet representation, and $G_U$, $G_{\Psi}$, and $B\theta$ could all be non-zero in general, quantization of $\text{Tr}[G_U]$ is expected from quantization of $\text{Tr}[G_{\Psi}]$ and $\text{Tr}[B\theta]$. $\text{Tr}[G_{\Psi}] = N(k+1)$, so if $\text{Tr}[G_U]$ happens to be trivial in a particular instance, $\text{Tr}[B\theta] = NB\theta$, so $B\theta = k+1$. Therefore, if $\text{Tr}[G_U]$ is non-zero but $G_U$ must yield the singlet representation in combination with $G_{\Psi}$ and $B\theta$, when the trace of either is quantized in multiples of $k+1$, then non-trivial $\text{Tr}[G_U]$ must also be quantized in multiples of $k+1$. We also note that, by incompressibility, bulk excitations encoded by $G_U$ are equivalent to boundary excitations encoded by $G_{\Psi}$~\cite{polychronakos2000noncommutative}. This is consistent with the fusion rules yielding quasi-hole or quasi-particle excitations with charge of $1/(k+1)$. 

The fusion rules are not expected necessarily to correspond to $SU(N)_{k}$ as for commutative, non-Abelian CS theory taking into account the finite matrix dimension and partial filling of the fuzzy sphere, which is expected to reduce symmetry and yield fuzzy coset space structure~\cite{aschieri2006dynamical, morariu_fractional_2005}. Identifying the appropriate representation theory is beyond the scope of the present work.  We expect results at $N \rightarrow \infty$ of CSMs to be relevant in determining the fusion rules, in particular their identification with the $W_{1+\infty}$ algebra ~\cite{Caracciolo:1995uj,PINZUL_2003}.

Taking into account that the Poisson bracket should be converted to a Lie derivative for a quantum theory defined over a fuzzy coset space more generally, or conventional conversion of the Poisson bracket to a commutator should be supplemented by additional statements, such as a statement of flat connection, taking into account generalization of partial derivatives to Lie derivatives on fuzzy coset spaces, we reformulate $2BR^2\left[z, z^{\dagger}\left(1+ z z^{\dagger} \right)^{-1} \right]$ to relate it to the topological invariant $\text{Tr}[C]$ introduced to study the QSkHE and related phases of matter~\cite{qskhe, patil2024effective, ay2024signatures, banerjee2024multiplicative}. We first set $2BR^2=1$ or equivalently absorb this prefactor into the $z$ and $z^{\dagger}$ matrices. We then reformulate the expression as
\begin{align}
\left[z, z^{\dagger}\left(1+ z z^{\dagger} \right)^{-1} \right] &= zz^{\dagger}\left(1+ z z^{\dagger} \right)^{-1} - z^{\dagger}\left(1+ z z^{\dagger} \right)^{-1} z  \nonumber \\
&= zz^{\dagger}\left(1+ z z^{\dagger} \right)^{-1} - z^{\dagger} z \left(1+ z^{\dagger}z \right)^{-1},
\end{align}
taking into account the quantum ordering conventions of Morariu and Polychronakos~\cite{morariu_fractional_2005}. We then redefine our expression as
\begin{align}
\left[z, z^{\dagger}\left(1+ z z^{\dagger} \right)^{-1} \right] &= zz^{\dagger}\left(1+ z z^{\dagger} \right)^{-1} - z^{\dagger} z \left(1+ z^{\dagger}z \right)^{-1}  \nonumber \\
&= \left(zz^{\dagger} - z^{\dagger} z \right) S^{occ}_{z, \dagger_{1,2}} \nonumber \\ 
&= \left[z,z^{\dagger}\right] S^{occ}_{z, \dagger_{1,2}},
\end{align}
where $S^{occ}_{z, \dagger_{1,2}}$ is $S^{occ}_{z, \dagger_{2}}=\left(1+ z z^{\dagger} \right)^{-1}$ if combined by matrix product with $zz^{\dagger}$, and it is $S^{occ}_{z, \dagger_{1}}=\left(1+ z^{\dagger} z \right)^{-1}$ if combined with $z^{\dagger}z$. That is, it is just notation to identify a shared factor of the terms obscured by the quantum ordering convention. We similarly identify the lowering/raising operators~\cite{morariu_fractional_2005} $z$ and $z^{\dagger}$ as $\left(S^{occ}_{\pm} \right)^{-1}$, yielding
\begin{align}
\left[z, z^{\dagger}\left(1+ z z^{\dagger} \right)^{-1} \right] &= \left[\left(S^{occ}_{-} \right)^{-1},\left(S^{occ}_{+} \right)^{-1}\right] S^{occ}_{z, \dagger_{1,2}}.
\label{ztosocc}
\end{align}
 We then identify the projected spin operator algebra of the QSkHE, in the context of the MCS theory on the sphere, as
\begin{align}
\left[S^{occ}_{-},S^{occ}_{+} \right] &= \left(C^{occ}_{-+z} \right)^{-1}S^{occ}_{z, \dagger_{1,2}},
\end{align}
where $\left(C^{occ}_{-+z} \right)^{-1}$ is the matrix inverse of $C^{occ}_{abc}$ in Eq.~\ref{flatconn} with $a=-$, $b=+$, $c=z$. The generalized quantization yielding $\text{Tr}\left[C^{occ}_{abc} \right] = k+1$ yields $\text{Tr}\left[\left(C^{occ}_{abc}\right)^{-1} \right] = 1/(k+1) = 1/3$ for $k=2$.

The quantization arguments of Polychronakos for MCS theory, generalized to account for the differential geometry of the partially-filled fuzzy sphere and related fusion rules, then also apply to quantization of $\text{Tr}[C]$. In particular, they support observed quantization of $\text{Tr}[C]$ to rational numbers in numerics for the multiplicative Chern insulator of the QSkHE. The $1/3$ contribution to $\text{Tr}[C]$ observed for individual unit cells as part of a generalized $4 \pi$ AB effect, within the framework of MCS theory, is consistent with  a bounded MCS theory for $N=4$ electrons at filling fraction $1/(k+1)$ and $k=2$.

In the next section, we apply calculations of filling fraction within finite MCS theory of Polychronakos to provide further insight into the topological response signature of the multiplicative Chern insulator also consistent with realization of $N=4$ MCS droplets at $k=2$. In later sections, we furthermore construct generalized gauge theories as $D$-dimensional arrays of small $N$ MCS droplets, to find that $k=2$ is also consistent with the $4\pi$ AB effect of the full MCI model, in addition to the numerics for individual $N=4$ unit cells of the lattice model.

\subsection{\texorpdfstring{$1/3$ for $N=4$}{1/3 for N=4} specifically and for merging of two \texorpdfstring{$N=2$}{N=2} parents of the MCI}

Here, we consider possible filling fractions specifically for $N=2$ and $N=4$ MCS droplets with partial filling and how they might yield $k=2$ for $N=4$ in the MCI.

The filling fraction for the MCS theory is $N/L$, where $N$ is the number of electrons and equal to the matrix dimension of the MCS theory, and $L=m+1$ is related to the monopole charge $m$. Filling fractions for the MCS theory on the fuzzy sphere must satisfy $(N-1)(k+1) \leq m$.

Each parent of the multiplicative Chern insulator has Chern number $\pm 1$, with matrix representation $N=2$. We can ask whether this is consistent with $1/3$ filling fraction for an $N=4$ fuzzy sphere if the two $N=2$ parent fuzzy spheres merge during the flux insertion process, to form an $N=4$ fuzzy sphere. This analysis of possible merging of partially-filled fuzzy spheres is summarized schematically in Fig.~\ref{fillfrac_merge}.

Chern number of $\pm1$ corresponds to $k=1$ and a $k+1=2$ \textit{bounded} MCS theory of Polychronakos. For $N=2$, $N/L = 1/2$ holds for $m=L-1 = 4-1 = 3$, which is greater than $(N-1)(k+1) = 2$. This corresponds to a difference in monopole strength compared to the maximum for this $L$, $\Delta_{N=2} m$, of $1$. That is, each $N=2$ droplet has a single bare state, as shown in Fig.~\ref{fillfrac_merge}.

 If the two $N=2$ fuzzy spheres merge, we realize an $N=4$ fuzzy sphere. For $N=4$, $N/L = 1/3$ holds for $m=L-1 = 12-1 = 11$, which is greater than $(N-1)(k+1) =9$. This corresponds to a difference in monopole strength compared to the maximum for this $L$, $\Delta_{N=4} m$, of $2$. A bounded/finite MCS theory of Polychronakos, of $k+1=3$, further corresponds to an unbounded MCS theory of Susskind with $k=2$, corresponding to filling fraction $\nu=1/2$, which is consistent with the $4 \pi$ AB effect of the MCI.

 Furthermore, $\Delta_{N=4} m = \Delta_{N=2} m + \Delta_{N=2} m$ reflects a conservation of bare flux quanta. However, if a flux quantum is inserted through the MCS sphere/CSM ring, the particle at highest quasi-momentum can be peeled from the surface of the Fermi sea, increasing its quasi-momentum by $1$. Such an excitation can create a quasi-hole of charge $1/(k+1)$, as mentioned in~\cite{polychronakos2001quantumcyl}, which is also shown in Fig.~\ref{fillfrac_merge} as part of the merging process specifically due to flux insertion.

 From this perspective, $\text{Tr}\left[C\right]=1/3$ is consistent both with an $N=4$ MCS theory at $k=2$, but also with realization of an $N=4$ MCS droplet from merging of two $N=2$ MCS droplets, each at $k=1$, corresponding to the two parents of the multiplicative MCI considered in numerics~\cite{banerjee2024multiplicative}. The flux insertion process breaks the tensor product structure of the bulk Hamiltonian, consistent with a notion of coupling and merging of the parent fuzzy spheres.

\begin{figure}[t]
    \centering
  \includegraphics[width=\columnwidth]{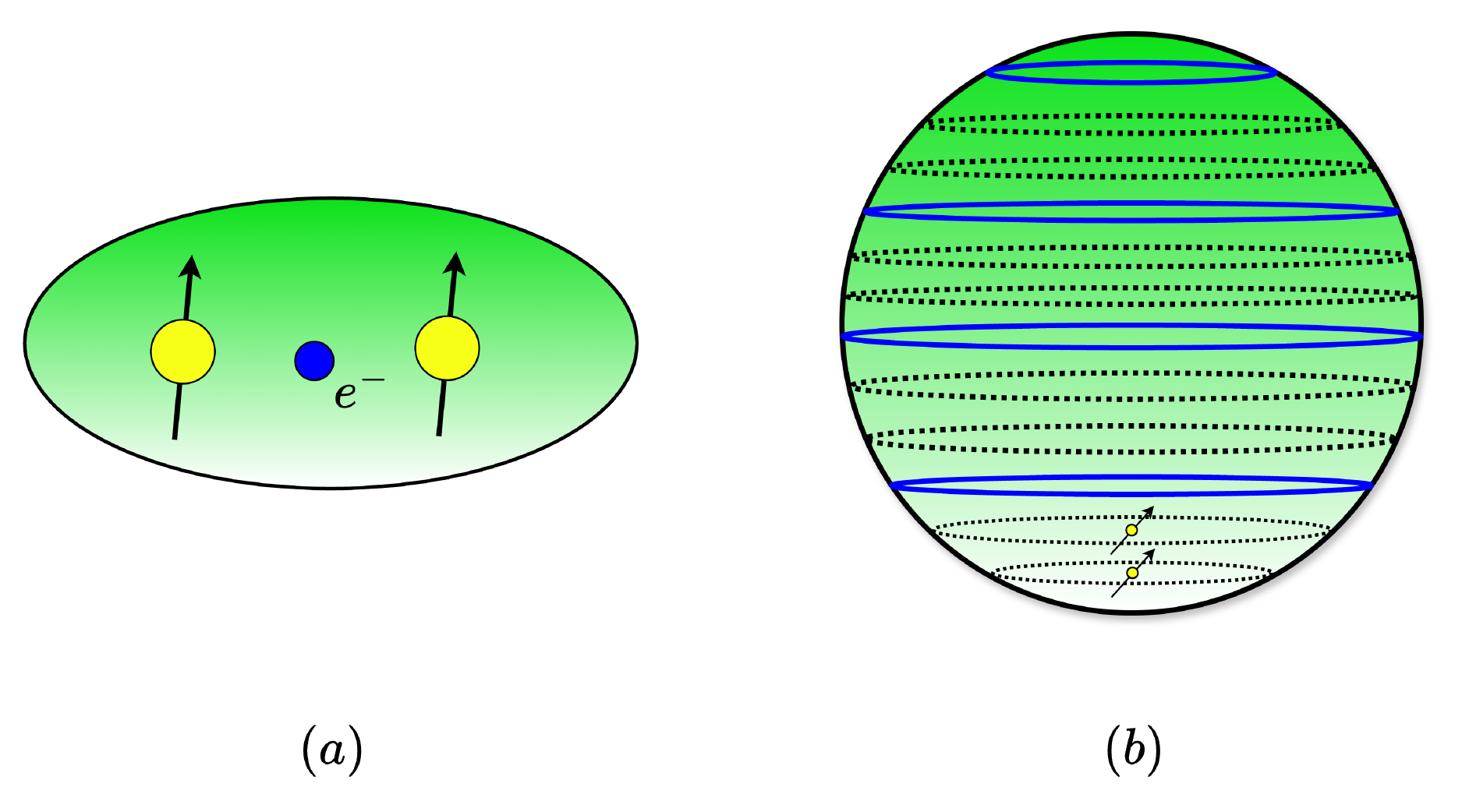}
  \caption{(a) Composite particle with two flux quanta and an electron $e^-$ for FQHE plateau at filling fraction $\nu=1/2$. (b) Generalization of the composite particle depicted in (a) to an $N=4$ MCS theory, with QH droplet identified with the $e^-$ of (a) and two bare levels near the south pole identified with the two flux quanta in (a).}
  \label{genJainparticle}
\end{figure}

We note that the number of bare states in the $N=2$ and $N=4$ MCS droplet models, $k=1$ and $k=2$, respectively, is consistent with interpretation of the small $N$ MCS droplet as a generalization of Jain's composite particle for related Laughlin states~\cite{jain1994}, if the droplet on the sphere is identified as the charge of a Jain particle combined with $k$ flux quanta, as shown in Fig.~\ref{genJainparticle}. This interpretation is also consistent with a physically meaningful interpretation of filling fraction $N/L$ both for the case of $N \rightarrow \infty$, corresponding to a large, bounded MCS droplet, but also for the case of $N \rightarrow 1$, which may be interpreted as effectively integrating out internal degrees of freedom of the MCS droplet for partial filling of the sphere, reducing the droplet to an effective unit point charge. The number of bare flux quanta, $\Delta_{N} m$, may be taken as $k$. In the former case ($N \rightarrow \infty$)
\begin{align}
\frac{N}{L} = \frac{N}{(N-1)(k+1) + k} \rightarrow \frac{1}{k+1}.
\end{align}

In the latter case, in which $N \rightarrow 1$,
\begin{align}
\frac{N}{L} = \frac{N}{(N-1)(k+1) + k} \rightarrow \frac{1}{k},
\end{align}
in agreement with reduction of the MCS droplet to a Jain composite particle with unit charge and $k$ flux quanta, the composite particle of the $\nu=1/k$ FQH plateau~\cite{jain1994}.

 From this perspective, we can interpret the contribution to $\text{Tr}[C]$ of $1/3$ as a signature of the filling fraction of individual MCS $N=4$ droplets of $1/(k+1) = 1/3$, or $k=2$. The occurrence of this signature in combination with a $4 \pi$ AB effect of the MCI is then consistent with the full MCI system described by the counterpart unbounded MCS theory at level $k=2$, or filling fraction $1/2$, yet consisting of multiple $N=4$ MCS droplets, each at level $k+1=3$. In the next section, we will strengthen this link by explicitly constructing a Yang-Mills action from the bounded MCS theory.

\section{Tiling construction of generalized 2+1 D U(N) Yang-Mills theory from finite N matrix Chern-Simons theory}\label{tiling}

In the previous section, we considered a single $N=4$ MCS/CSM droplet, or pair of $N=2$ MCS/CSM droplets, to model a single $4 \times 4$ unit cell of the MCI. While this captures some of the local signatures observed in numerics, namely the non-trivial topological invariant $\textrm{Tr}[C]$~\cite{banerjee2024multiplicative, patil2024effective}, understanding other aspects of the topological response of the MCI requires modeling of multiple unit cells, rather than a single unit cell. This naturally leads us to explore formulations of \textit{arrays} of many MCS/CSM droplets, each droplet for small matrix dimension $N$. Two possibilities are to 1. construct an array from many identical MCS/CSM droplets, yielding a theory very similar to a Bloch Hamiltonian description in possessing analogues of crystalline point group symmetries, and 2. construct an array from many MCS/CSM droplets, which may be distinct from one another, which might model e.g., disorder or local perturbations such as the local insertion of magnetic flux. In this section, we consider the higher-symmetry constructions of point 1: we first model the MCI by constructing a $2+1$ D Yang-Mills theory similarly to the construction of a $1+1$ D Yang-Mills theory from a finite $N$ matrix model by Polychronakos~\cite{polychronakos2001quantumcyl}. We construct the $2+1$ D Yang-Mills theory from the $N=4$ MCS theory for $1/(k+1) = 1/3$ relevant to numerics~\cite{banerjee2024multiplicative}. However, we enforce spherical rather than cylindrical constraints and later apply concepts of extra fuzzy dimensions for deeper understanding from the perspective of the QSkHE. We note that the following construction is analogous to construction of a Bloch Hamiltonian with $N\times N$ matrix representation defined over a two-torus Brillouin zone, starting from an $N\times N$ Hamiltonian for a single unit cell.

We begin by demanding that the state be the same on all copies of the finite $N=4$ MCS theory. This corresponds to the condition that fields in different copies are gauge equivalent, meaning the generator shifting copies in either spatial direction introduced by tiling, $U_1$ and $U_2$, is a unitary gauge transformation.
\begin{align}\label{conds}
&U_1X_1U_1^{-1} = X_1 + 2 \pi R \\
&U_2X_2U_2^{-1} = X_2 + 2 \pi R \\
&U_{1,2} A_o U_{1,2}^{-1} = A_o \\
&U_{1,2} \Psi = \exp{i\alpha_{1,2}} \Psi,
\end{align}
with $\boldsymbol{\alpha} = \left(\alpha_1, \alpha_2 \right)$.
This can be explicitly realized by parametrizing indices $I,J$ of $(X_a)_{IJ},(A_0)_{IJ},\Psi_I$ as:\\
\begin{align}
    I &= i + nN, \hspace{0.5cm} i=1,...,N, \hspace{0.5cm} n=....,-1,0,1,... \\
     J &= j + mM, \hspace{0.5cm} j=1,...,M, \hspace{0.5cm} m=....,-1,0,1,... 
\end{align}
 This corresponds to splitting the matrix field $X_a$ in terms of $N \times N$ blocks. The diagonal block labeled by $n$ and $m$ encodes $N$ electrons in the $(n,m)$ copy of the covering space, while off-diagonal blocks, labeled by $(\eta, \mu)$, encode effective interactions between the $(n,m)$ block and the $(n',m')$ block. By periodicity, the $\left(n,m \right)$ copy is the same as the $\left(0,0\right)$ copy, save that it is shifted by $2\pi R n$ in $X_1$ and by $2\pi R m$ in $X_2$. 

Interactions between $(n,m)$ and $(n',m')$ copies must depend only on the distance between these copies, or $|(n,m)-(n',m')|$, in effect. The Lagrange multiplier $\left(A_o \right)_{m,n}$ must similarly impose constraints on the $(n,m)$ and $(n',m')$ copies, which only depends on the distance $|(n,m)-(n',m')|$. In addition, $\Psi_{nm}$ represents the boundary of the state in copy $(n,m)$, and must therefore be the same for all copies up to an irrelevant phase. Labeling the $(n,m)$ copy by $\boldsymbol{r}_{nm}$, we may therefore write 
\begin{align}
&\left(X_1\right)_{\boldsymbol{r}_{nm},\boldsymbol{r}_{n'm'}} = \left(X_1\right)_{|\boldsymbol{r}_{nm}-\boldsymbol{r}_{n'm'}|} +2 \pi R n \delta_{n n'}, \\
&\left(X_2\right)_{\boldsymbol{r}_{nm},\boldsymbol{r}_{n'm'}} = \left(X_2\right)_{|\boldsymbol{r}_{nm}-\boldsymbol{r}_{n'm'}|} +2 \pi R m \delta_{mm'}, \\
&\left(A_o\right)_{\boldsymbol{r}_{nm},\boldsymbol{r}_{n'm'}} = \left(A_o\right)_{|\boldsymbol{r}_{nm}-\boldsymbol{r}_{n'm'}|}, \\
&\Psi_{\boldsymbol{r}_{nm}} = \exp{(i \boldsymbol{r}_{nm}\cdot \boldsymbol{\alpha})} \Psi.
\end{align}

$U_1$ ($U_2$) is then a shift to a nearest-neighbor copy (effectively a shift by one unit) in the $I$ ($J$) index/direction. \\

We furthermore have 
\begin{align}
    \left(X_a\right)_{\boldsymbol{r}_{nm}} &= \left(X^{\dagger}_a\right)_{-\boldsymbol{r}_{nm}}, \\
    \left(A_o\right)_{\boldsymbol{r}_{nm}} &= \left(A^{\dagger}_o\right)_{-\boldsymbol{r}_{nm}}.
\end{align}

We therefore define the Fourier transformations in terms of two-component variable $\boldsymbol{\sigma}=\left(\sigma_I, \sigma_J \right)$ and each of $\sigma_I$ and $\sigma_J$ $2\pi$-periodic similarly to momentum components of a two-torus Brillouin zone with all lattice constants set to $1$. This interpretation is especially relevant to modeling of the MCI. These Fourier transformations are then

\begin{align}
X_a \left( \boldsymbol{\sigma} \right) &= \sum_{\boldsymbol{r}_{nm}} \left( X_a\right)_{\boldsymbol{r}_{nm}} \exp{\left(i \boldsymbol{r}_{nm} \cdot \boldsymbol{\sigma}\right)}, \\
A_o \left( \boldsymbol{\sigma} \right) &= \sum_{\boldsymbol{r}_{nm}} \left( A_o\right)_{\boldsymbol{r}_{nm}} \exp{\left(i \boldsymbol{r}_{nm} \cdot \boldsymbol{\sigma}\right)}, \\
\Psi\left( \boldsymbol{\sigma} \right) &= \sum_{\boldsymbol{r}_{nm}} \Psi_{\boldsymbol{r}_{nm}} \exp{\left(i \boldsymbol{r}_{nm} \cdot \boldsymbol{\sigma}\right)}.
\end{align}

Matrix multiplication in $I,J$ indices translates to pointwise matrix multiplication in $\boldsymbol{\sigma}$. $I$/$J$-trace becomes $\boldsymbol{\sigma}$ integration and matrix trace. 

Using these results, we transition from the finite matrix model for the sphere in terms of chord length matrices $w$ and $w^{\dagger}$ as in Eq.~\ref{lchord} with additional harmonic potential term $V(w^{\dagger}w)=\frac{1}{2}w^2\text{Tr}[w^{\dagger}w]$~\cite{morariu_fractional_2005},
\begin{align}\label{suth}
    \mathcal{L}(w,w^{\dagger})=&i\frac{B}{2}\text{Tr}\left[w^{\dagger}\dot{w}-iA_0([w,w^{\dagger}]-2\theta)\right]\nonumber\\
    &\qquad+V(w^{\dagger}w)+\Psi^{\dagger}(i\dot{\Psi}+A_0\Psi),
\end{align}
to an action constructed as an array of identical $N\times N$ finite matrix model copies similarly to the approach of Polychronakos for constructing a $1+1$ D Yang-Mills theory~\cite{polychronakos2001quantumcyl}. In terms of one-form components $\Omega$ of matrix dimension $N$, the Lagrangian may be written compactly as
\begin{align}
S =\int dt d\boldsymbol{\sigma} i (B/2) \text{Tr}[  &\Omega d\Omega - i \Omega_o \left(\Omega \Omega-2\theta \right) -  \partial_{\sigma} \Omega_o]\nonumber\\
&+ \Psi^{\dagger} \left( i \dot{\Psi} + \Omega_o \Psi\right).
\end{align}

However, in this formulation similar to that of Polychronakos~\cite{polychronakos2001quantumcyl}, the action naively appears to be a $2+1$ D $U(N)$ Yang-Mills action based on the number of Cartesian spatial coordinates encoded by $\boldsymbol{\sigma}$ in addition to the time dimension. However, the matrix structure alone still encodes two spatial dimensions in fuzzy space coordinates~\cite{aschieri2006dynamical}. This is the case even for $N$ small within the framework of the QSkHE~\cite{qskhe,patil2024effective}. That these matrices still encode two spatial dimensions is actually also consistent with MCS theory for general finite $N$, although small $N$ was not explicitly considered. We therefore make this dimensionality explicit by rewriting the action in the framework of extra fuzzy dimensions~\cite{aschieri2004dimensional, aschieri2007, Aschieri:2004vh, aschieri2006dynamical}, assuming a global product structure due to the construction from multiple copies of the finite $N$ MCS theory. Taking into account the isomorphism $U(N) \cong U(1) \times SU(N)/\mathbb{Z}_N$ ~\cite{Tong:2016kpv}, we first separate each $U(N)$ gauge field component into a $U(1)$ component for a Cartesian coordinate, $A$, and an $SU(N)/\mathbb{Z}_N$ component for the fuzzy coset space coordinates, $\hat{F}$. That is, we take $\Omega d \Omega = AdA \wedge \hat{F}$. We decompose $\Psi$ similarly as $\Psi = \psi \otimes \hat{\Psi}$. However, although it is natural within commutative, non-Abelian CS theory to treat the matrix part as $SU(N)/\mathbb{Z}_N$, the matrix theory here is potentially for a \textit{partially-filled} sphere, and therefore its algebraic structure may include deviations from $SU(N)/\mathbb{Z}_N$ as discussed in the previous section. The action is then
\begin{align}
S &= \int dt d\boldsymbol{\sigma} i (B/2) \text{Tr}[  AdA \wedge \hat{F} - i A_o \left(A A-2\theta \right)\wedge \hat{F}  \nonumber\\
&-\partial_{\sigma}A_o \wedge \hat{F}] + \psi^{\dagger}\otimes \hat{\Psi}^{\dagger} \left( i \dot{\psi}\otimes \dot{\hat{\Psi}} + A_o \wedge\hat{F} \psi \otimes \hat{\Psi}\right)
\end{align}

At this level of treatment, we may also convert a $Q=2$ prefactor in front of a term such as $AdA \wedge \hat{F}$ into a $1/Q$ prefactor, by first merging the $\hat{F}$ with $A$ to work in terms of $\Omega$ gauge field components for calculational purposes. Such terms are then quadratic in $\Omega$ components, and may therefore naively be replaced using an EOM to convert the prefactor $Q$ to $1/Q$~\cite{Tong:2016kpv}. The resultant $U(N)$ gauge field components, now with prefactor $1/Q$, may then be formulated in the framework of extra fuzzy dimensions again, in terms of $U(1)$ component $a$ and $SU(N)/\mathbb{Z}_N$ component (with deformations) $\hat{f}$ gauge fields. This illustrates how the $4\pi$ AB effect of the multiplicative Chern insulator is identified with a generalization of the $\nu=1/2$ FQH state despite realization in an effectively non-interacting lattice tight-binding model via the effectively higher-dimensional term attributed to minimal realization of the QSkHE, of the form $ada \wedge \hat{f}$.

In this construction, the boundary term of the MCS theory, which ultimately yields the level shift of the CS level to $k+1$, is reformulated as the source term for charged particles in an unbounded MCS theory at CS level $k$. This tiling construction is then consistent with the MCI exhibiting signatures of a generalized $\nu=1/2$ FQH plateau globally, in combination with $\text{Tr}[C]$ including a $1/3$ contribution for individual unit cells, corresponding to a $\nu=1/3$ plateau for the corresponding bounded MCS theory. From this perspective, we can identify the quantum skyrmions of the QSkHE as low-lying charged excitations of the finite $N$---particularly small $N$---MCS theory/CSM, such as the quasi-holes/particles discussed by Polychronakos, in direct analogy to the commutative skyrmions that occur as low-lying charged excitations in the QHFM~\cite{PhysRevB.47.16419}. Although these skyrmions at small $N$ do not possess the smooth topological texture over a vector field defining the commutative skyrmion, they are charged objects formulated in terms of spin, and regain such a smooth texture when taken to the commutative limit. While these quantum skyrmions can occur in isolated MCS/CSM droplets, they can also occur in coupled arrays of droplets, in which case they play the role in an effectively higher-dimensional QSkHE that charged particles play in the QHE. This is directly analogous also to the role played by $D$-branes ($D=2$ being the most directly relevant case) in higher-dimensional QHEs, which respond to effective fields similarly to charged point particles in the regime of thin branes~\cite{Bernevig:2002eq, Zhang:2001xs}.

We finally point out an important feature of this tiling construction of Polychronakos, through an example of its application within the context of the QSkHE. Morariu and Polychronakos~\cite{morariu_fractional_2005} note that a harmonic oscillator potential on the plane, $\tilde{V}$, maps to a constant electric field in the vertical direction on the fuzzy sphere. From the Yang-Mills theory constructed here in terms of finite $N$ MCS droplets, we may furthermore see the harmonic oscillator potential term in the finite $N$ MCS theory, for each bounded droplet, serves as an effective Zeeman field term such as in the Lagrangian for a quantum Hall ferromagnet (QHFM)~\cite{PhysRevB.47.16419}. (More specifically, this harmonic potential term takes a form similar to the Zeeman field term $B\bar{\Psi}\sigma_z \Psi$ in work on the QHFM, for instance, by Sondhi~\emph{et al.}~\cite{PhysRevB.47.16419}.) We note that these potentials may be useful in defining exchange couplings between neighboring spins in more complex constructions, a point to be investigated in future work. 

In this scenario, we see that a combination of Zeeman and orbital magnetic fields in a Lagrangian with extra fuzzy dimensions is equivalent to a combination of applied electric and orbital magnetic fields in a theory without extra fuzzy dimensions. This interpretation and mapping is also consistent with observed topological response signatures in numerics for the Bernevig-Hughes-Zhang model associated with the dimensionally-reduced 4+1 D Chern insulator state~\cite{ay2024signatures}. This further illustrates the significance of the tiling construction of Polychronakos, if interpreted within the framework of the QSkHE: it translates between physics of commutative, non-Abelian quantum field theory and microscopic field theories of the QSkHE. Related to this point, the tiling construction outlined here also illustrates that non-Abelian FQHE states are potentially more deeply understood within the framework of the QSkHE, which reveals possible deformations in algebraic structure associated with finite $N$ MCS droplets effectively encoded in the non-Abelian gauge fields. QH bilayer systems, in particular, should be useful platforms for observation of the QSkHE~\cite{girvin1996multicomponent}.

\section{Modeling the quantum skyrmion Hall effect within matrix Chern-Simons theory and anisotropic fuzzification}\label{array}

\subsection{Two-dimensional array of small \texorpdfstring{$N$}{N} matrix Chern-Simons droplets}

To model lower-symmetry microscopic field theories of the QSkHE than possible with the tiling construction to study e.g., disorder or effects of local perturbations, in this section we introduce minimal formulations of possible actions and Hamiltonians for the MCI and related states of the QSkHE within the framework of anisotropic fuzzification introduced in Patil et al.~\cite{patil2024effective}. We simply illustrate constructions within this framework, postponing characterization given the potential complexity of this analysis. 

We will construct such models as arrays of spatially-separated small $N$ MCS droplets or corresponding CSMs, with some coupling between droplets. For this purpose, we re-state the Lagrange density of the finite matrix action for the small $N$ MCS droplet~\cite{morariu_fractional_2005}, which will be the Lagrange density of the $i$\textsuperscript{th} droplet in the array, $\mathcal{L}^{(i)}$. $\mathcal{L}^{(i)}$ characterizes both a 2+1 D MCS droplet and an equivalent 1+1 D CSM as discussed earlier in section~\ref{secrew}  in terms of $N \times N$ matrix gauge fields $\omega_{(i)}$ and $\omega^{\dagger}_{(i)}$, as
\begin{align}
    \mathcal{L}^{(i)}=&i\frac{B^{(i)}}{2}\text{Tr}\left[w^{\dagger}_{(i)}\dot{w}^{}_{(i)}-iA_{0,_{(i)}}\left([w^{}_{(i)},w^{\dagger}_{(i)}]-2\theta^{(i)}\right)\right]\nonumber\\
    &+V^{(i)}(w^{\dagger}_{(i)}w^{}_{(i)})+\Psi^{\dagger}_{(i)}\left(i\dot{\Psi}_{(i)}+A_{0,{(i)}}\Psi_{(i)}\right)
\end{align}
For ease of notation, we suppress the $(i)$ labeling for all parameters within the expression for the $i$\textsuperscript{th} Lagrangian, but stress that we consider scenarios in this section, in which the parameter set of $\mathcal{L}^{(i)}$ may differ from that of $\mathcal{L}^{(j)}$ ($i\neq j$), such that the tiling construction of Polychronakos as in the previous section is not necessarily possible due to low symmetry. We will then consider the scenario where $N$ is small relative to the possible values of $M$, where $i \in \{1, ..., M\}$.

Equivalently, it is useful to consider the counterpart CSM Hamiltonian for the $i$\textsuperscript{th} droplet, as 
\begin{equation}
H^{(i)}=\sum_{n=1}^N\frac{\omega_{(i)}}{2B_{(i)}}\left(p^{(i)}_n\right)^2+\sum_{n\neq m}\frac{k_{(i)}\left(k_{(i)}+1\right)}{4\sin^2\left[(\phi^{(i)}_n-\phi^{(i)}_m)/2\right]}
\label{iSuthHam}
\end{equation}
with $(\phi^{(i)}_n,p^{(i)}_n)$ being the co-ordinates and momenta of particles on the circle representing the $i$\textsuperscript{th} CSM.~\cite{morariu_fractional_2005}.

Having defined the Lagrangian and Hamiltonian for an individual MCS droplet in terms of its corresponding CSM, we will proceed in defining a Hamiltonian and Lagrangian for a collection of MCS droplets within anisotropic fuzzification as follows. We will first construct such an array in a Hamiltonian formulation, in terms of the CSM Hamiltonian. We will then define a  center for the $i$\textsuperscript{th} droplet within Lagrangian formulation. We will then construct $D$-dimensional arrays of $M$ such droplets in a Lagrangian formulation, in terms of multiple instances of the MCS/CSM Lagrangian, by spatially separating them. The dimension of the array, $D$, is arbitrary and depends on how droplet centers are shifted and the nature of couplings between droplets: $D=1$ corresponds to a chain of droplets, a construction we will argue can model spin chains. $D=2$ corresponds to an array of droplets relevant to modeling 2D spin lattice models generalized within the QSkHE. Within this formulation, we will be able to connect these droplet arrays to treatments of multiple fuzzy spheres in the literature, in which multiple fuzzy spheres may merge. With this formulation of merging of fuzzy spheres, we will comment on how the MCI generalizes from the commutative CS theory of the FQHE, and how flux insertion may induce merging of a pair of $N=2$ parent droplets into a single $N=4$ child droplet, defining a previously-unidentified pairing mechanism of lattice tight-binding models.

\subsection{Array of matrix Chern-Simons droplets within Hamiltonian formulation of Calogero-Sutherland model}

In this subsection, we construct Hamiltonians for the QSkHE as a $D$-dimensional array of coupled CSMs in a Hamiltonian formulation as harmonic oscillators for $N$ electrons and fractional exclusion statistics related to the level $k+1$ of the corresponding MCS theory, using the Hamiltonian Eq.~\ref{suthHam}, formulated specifically as in Eq.~\ref{iSuthHam}. Such a formulation provides insight into the effective meaning of the later Lagrangian formulation of the MCI within the framework of anisotropic fuzzification. 

In this minimal description, the electron number for each droplet is conserved, with the possibility of changes in electron number for individual droplets to be explored in future work. We note that an effective current of quantum skyrmions (quasi-hole/particle excitations) could still be possible, in principle, even if electrons themselves are confined to individual droplets. A minimal coupling between different finite $N$ MCS/CSM droplets is modeled after the interaction term for each individual droplet, but taken to be between nearest neighbors, denoted by $\langle i,j \rangle$. A nearest-neighbor coupling $k_{(ij)}$ is introduced as well, such that the full Hamiltonian is \\
\begin{align}
H&=\sum_{i=1}^M H^{(i)} + \sum_{\langle i,j\rangle}\sum^N_{n}\sum^N_{m}\frac{k_{(ij)}\left(k_{(ij)}+1\right)}{4\sin^2\left[(\phi^{(i)}_n-\phi^{(j)}_m)/2\right]} .
\end{align}

For sufficiently strong coupling between droplets, it is reasonable to assume one effective limit of the array of droplets within the Hamiltonian description as a single CSM of matrix dimension $NM$ with uniform couplings, or
\begin{equation}
H=\sum_{n=1}^{NM}\frac{\omega}{2B}p_n^2+\sum_{n\neq m}\frac{k\left(k+1\right)}{4\sin^2\left[(\phi_n-\phi_m)/2\right]}.
\end{equation}

In contrast to the tiling construction of section~\ref{tiling}, this limit, corresponding to merging of the $M$ coupled small $N$ MCS droplets/Sutherland models into a single large, homogeneous MCS/CSM droplet for $NM$ electrons constitutes reduction of an anisotropically-fuzzified, higher-dimensional, potentially-deformed fuzzy sphere, specifically the deformed fuzzy four-sphere $S^4_{(T)F}$ as defined in Patil~\emph{et al.}~\cite{patil2024effective}, to a lower-dimensional, potentially-deformed fuzzy sphere, specifically the deformed fuzzy two-sphere $S^2_{(T)F}$. We note that taking this limit is then a dimensional reduction within available machinery. In this limit, electrons are free to move through the entire system, also.

\subsection{Constructing lower-symmetry \texorpdfstring{$D$}{D}-dimensional arrays of 
spatially-separated finite \texorpdfstring{$N$}{N} matrix Chern-Simons descriptions of 
quantum Hall droplets}

\begin{figure}[tbp]
    \centering
  \includegraphics[width=\columnwidth]{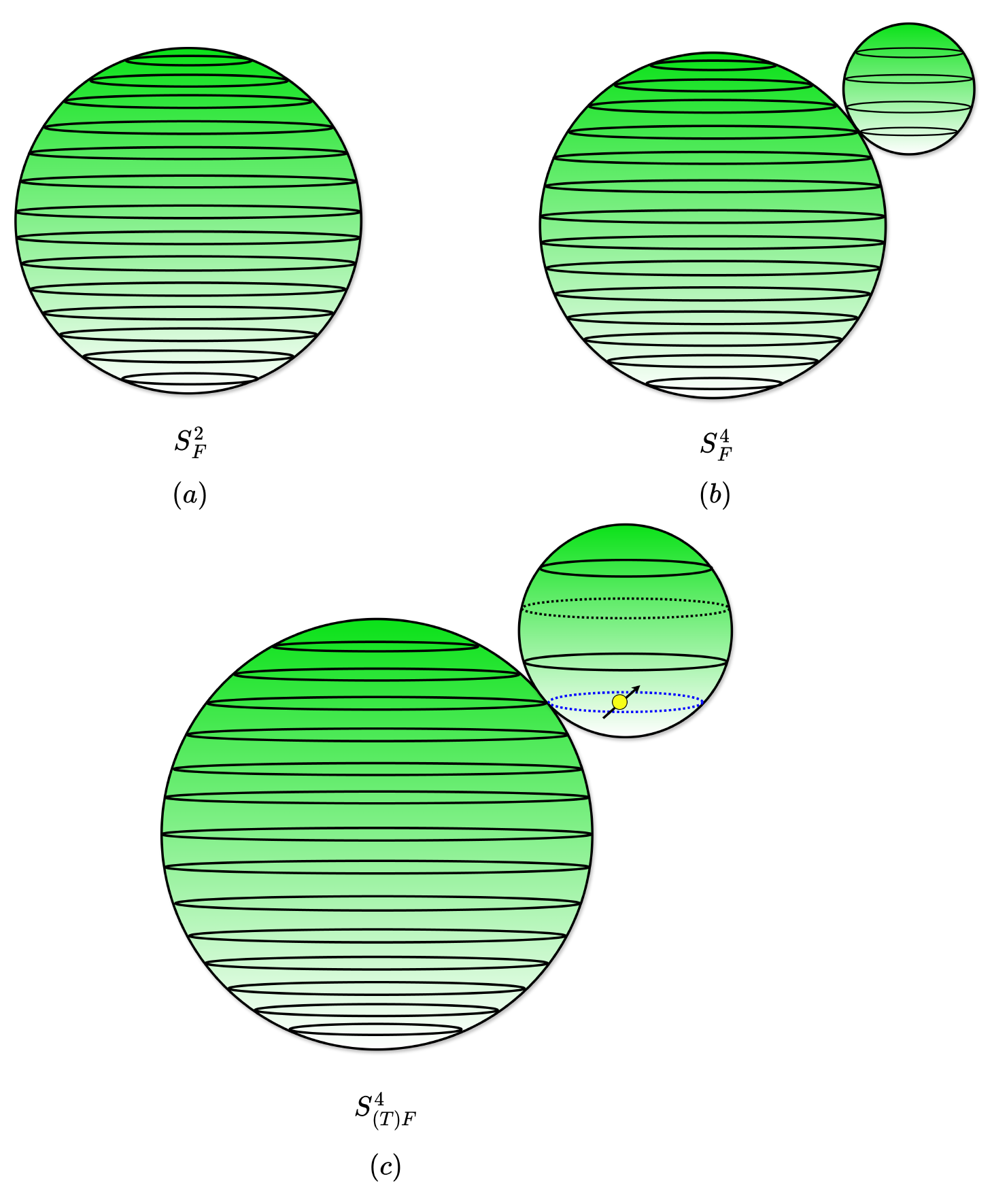}
  \caption{(a) Schematic representation of a fuzzy two-sphere $S^2_F$, with the full two-sphere approximated as $M$ orbitals of a LLL. (b) Schematic representation of a fuzzy four-sphere $S^4_F$ as a nesting of fuzzy two-spheres $S^2_F$ at two levels. Each orbital/flux quantum of the $S^2_F$ in (a) may itself be generalized to a $S^2_F$~\cite{hasebe2004dimensional}, here represented by the smaller $S^2_F$ with four orbitals of a LLL. (c) Microscopic quantum field theory of the QSkHE defined over a partially-filled fuzzy four-sphere, or deformed fuzzy four-sphere $S^4_{(T)F}$. Partial filling is illustrated by partial filling of the smaller fuzzy two-sphere, yielding a deformed fuzzy two-sphere $S^2_{(T)F}$ with $N=2$ electrons and one bare flux quantum.}
  \label{partialfill_fuzz4sphere}
\end{figure}

In this section, we will define $M$ MCS/CSM droplets, with each MCS/CSM droplet defined by an $N \times N$ Lagrangian with a potentially unique parameter set (the $i$\textsuperscript{th} parameter set). We will then embed these $M$ potentially distinct MCS/CSM droplet Lagrangians into an $NM \times NM$ matrix on its block diagonal to construct a Lagrangian for an array of $M$ such MCS/CSM droplets. For $N=4$ and $M\gg N$, such a construction can model the entire MCI system with disorder or other terms breaking the discrete translational symmetries inherent to the tiling construction. We will later discuss the rich topology of such constructions, given that each MCS/CSM droplet with $N$ electrons could have distinct topology, yielding topological states in the full  system of $M$ such MCS/CSM droplets far more general than that possible in the tiling construction.

While this embedding of $M$ droplets, each of matrix dimension $N$, combined with suitable couplings, may define a $D$-dimensional array of spatially-separated droplets in effect, we will also explicitly define a center for each droplet. To do so, we first review related work, such that we can then employ its approach. Bal~\emph{et al.}~\cite{Bal:2001cs} consider such a construction, although there is a distinction between their construction and the present one, in that they consider matrix dimension $N$ to be the orbital degeneracy of the LLL, rather than the number of electrons as in the present work. While this may be important for topological characterization and classification, it does not affect definition of droplet centers and the $D$-dimensional array of droplets, as we show.

In~\cite{Bal:2001cs}, multiple fuzzy spheres are defined in terms of the Lagrangian,
\begin{equation}
    L=-\text{Tr}V[X]
\end{equation}
with $V[X]$ containing terms for three-point and quartic  interactions between one-form matrix representations $X_{\mu}$. $X_{\mu}$ define $M$ fuzzy spheres and may therefore be expressed as a direct sum of $N \times N$ matrix one-forms, each defining an individual fuzzy sphere, or 
\begin{align}\label{multifuzzy1}
    X_{\mu}=\oplus^M_i X^{(i)}_{\mu},
\end{align}
with $X^{(i)}_{\mu}$ the matrix one-forms for the $i$\textsuperscript{th} fuzzy sphere.
The $i$\textsuperscript{th} fuzzy sphere may have a center and radius potentially specific to it, and the $i$\textsuperscript{th} center $\boldsymbol{R}_{(i)}$, with components $R^{(i)}_{\mu}$, and radius $\rho_{(i)}$ are given as,
 \begin{equation}\label{multifuzzy2}
     \quad [X^{(i)}_{\mu},X^{(i)}_{\nu}]=i\epsilon^{\mu\nu\rho}(X^{(i)}_{\rho}-R^{(i)}_{\rho}),\quad (X^{(i)}_{\mu}-R^{(i)}_{\mu})^2=\rho_{(i)}^2.
 \end{equation}

We apply the above construction, substituting the $N \times N$ Lagrangian $\mathcal{L}^{(i)}$ for $N$ electrons, for the $N \times N$ matrix $X^{(i)}$. To do so, we define a center for $\mathcal{L}^{(i)}$ as $\boldsymbol{R}_{(i)}$ according to the construction of Bal~\emph{et al.}~\cite{Bal:2001cs}, and define the $i$\textsuperscript{th} $N \times N$ Lagrangian for a MCS droplet with center $\boldsymbol{R}_{(i)}$ with components $R^{(i)}_{\mu}$, $\mathcal{L}^{(i)}(\boldsymbol{R}_{(i)})$, by generalizing the shift formulation in Bal~\emph{et al.}~\cite{Bal:2001cs} as

\begin{align}
\left[\omega_{\mu}^{(i)}, \omega_{\nu}^{(i)} \right] = C^{(i)}_{\mu \nu \rho} \left(\omega^{(i)}_{\rho} - R^{(i)}_{\rho
} \right),
\end{align}
where $C^{(i)}_{\mu \nu \rho}$ encodes the fusion rules of the MCS/CSM droplet, taking non-trivial values in the presence of quasi-hole/particle excitations of charge $1/(k+1)$ for partial filling of the $i$\textsuperscript{th} droplet.
We then express the Lagrangian for the array of $M$ droplets as
\begin{equation}\label{anisofuzz_fullL}
    \mathcal{L}=\text{Tr}_{MN}\left[\oplus_i^M \mathcal{L}^{(i)}(\boldsymbol{R}_{(i)})\right]=\oplus_i^M \text{Tr}_{N} \left[\mathcal{L}^{(i)}(\boldsymbol{R}_{(i)})\right],
\end{equation}
as the trace over the entire $MN \times MN$ matrix reduces to a trace on each $N \times N$ block with $\mathcal{L}(\boldsymbol{R})$ being the matrix functional in the MCS Lagrangian.

We may identify these Lagrangians describing an array of MCS droplets---to which we will next add inter-droplet coupling---as within the framework of anisotropic fuzzification introduced in Patil~\emph{et al.} as follows: we consider a lattice of $M$ flux quanta defining a fuzzy two-sphere $S^2_{F,M}$, and generalize each flux quantum itself to a fuzzy two-sphere of dimension $N$, or $S^2_{F,N}$, as done by Hasebe and Kimura~\cite{hasebe2004dimensional}, to construct a fuzzy four-sphere $S^4_F$ as a nesting of fuzzy two-spheres of type $S^2_{F,M}$ and $S^2_{F,N}$. We may then construct a MCS/CSM droplet with $N$ electrons, for each of the $M$ fuzzy spheres of type $S^2_{F,N}$, by partially filling each of the $S^2_{F,N}$ fuzzy spheres, corresponding to the Lagrangian $\mathcal{L}$. For $M$ distinct from $N$, this constitutes a Lagrangian within anisotropic fuzzification, and potential deformations from $S^4_F$ due to partial filling of the effective fuzzy four-sphere. For $M\gg N=4$, the anisotropic fuzzification is relevant to description of the MCI Hamiltonian written in real-space. This identification of the Lagrangians presented in this section with anisotropic fuzzification of Patil~\emph{et al.}~\cite{patil2024effective} is illustrated in Fig.~\ref{partialfill_fuzz4sphere}.

 To model the MCI, it is important to also discuss methods of coupling $N \times N$ blocks in $\mathcal{L}$. Rather than consider coupling for the more complex case of partial filling of fuzzy spheres, by generalizing from the Lagrangian for multiple partially-filled fuzzy spheres Eq.~\ref{anisofuzz_fullL} to include additional coupling between the partially-filled fuzzy spheres, we review a related, coarse-grained description of a QHE system given by a CS action for multiple fuzzy spheres and additional coupling~\cite{azuma2004nonperturbative}. This type of action and the phase structure, or which saddle point/fuzzy sphere configurations dominate and at what parameter value, as well as discussion specifically in terms of fuzzy branes, is furthermore discussed in a number of related works~\cite{alekseev2000brane, Hashimoto:2001xy,Bal:2001cs, jatkar2002matrix}. 
 
 Making use of the notation of Eq.~\ref{multifuzzy1} to streamline discussion, the action considered for studying coupling between multiple fuzzy spheres takes the form~\cite{azuma2004nonperturbative}
 \begin{equation}\label{braneact}
     S=N\text{Tr}\left[-\frac{1}{4}[X_{\mu},X_{\nu}]^2+\frac{2i\alpha}{3}\epsilon_{\mu\nu\rho}X^{\mu}X^{\nu}X^{\rho}\right],
 \end{equation}
 where $\alpha$ was taken to be a real parameter to facilitate Monte Carlo simulations. The CS term here is therefore real rather than imaginary as is standard, and $\alpha$ may furthermore take any real value without breaking symmetries in contrast to the standard CS term. We will examine related one-loop calculations of this work, which supported these Monte Carlo simulations, but specifically for scenarios relevant to modeling the MCI. 
 
 This action possesses saddle point configurations (phases) given by the EOM,
 \begin{equation}
     \left[X_{\nu},[X_{\nu}, X_{\mu}]+i\alpha \epsilon_{\mu\nu\rho}X_{\rho}\right]=0.
 \end{equation}
 The multi-fuzzy sphere configurations we described above in Eq. ~\ref{multifuzzy1} and ~\ref{multifuzzy2} are classical solutions to this fuzzy brane action Eq.~\ref{braneact}. These distinct solutions are characterized through the representations of $SU(2)$ in these works. We distinguish, in particular, between one large (e.g., $NM \times NM$ in the context of $\mathcal{L}$) irreducible representation of $SU(2)$ or the direct sum of reducible representations (e.g., $M$ blocks, each being $N \times N$ as in the expression for $\mathcal{L}$). The latter, reducible representation is known as a multi-fuzzy sphere configuration. 
 
In Jatkar~\emph{et al.}~\cite{jatkar2002matrix}, they considered the energy cost---and consequently stability---of different such (multi) fuzzy sphere configurations based on the the Lagrangian, with multiple saddle points potentially occurring at each set of parameter values and the dominant one minimizing the free energy. The stability of a configuration is therefore dictated by the perturbation around a given saddle-point configuration. Results of quantum Monte Carlo and supporting analytical calculations~\cite{azuma2004nonperturbative} find that off-diagonal elements of the non-linear terms couple different reducible sectors to induce phase transitions, in which the multiple blocks of the reducible saddle-point solution merge into a single, enlarged irreducible solution, for the case of co-incident spheres.

While such results are not directly relevant to the array of spatially-separated $N=4$ droplets, as they were for co-incident spheres, they are more relevant to understanding the process by which a pair of $N=2$ fuzzy spheres for each unit cell of the MCI merge into a single $N=4$ fuzzy sphere during time-reversal symmetric flux insertion, yielding non-trivial $\text{Tr}[C]$ signatures for flux $\phi$ in the vicinity of one flux quantum $\phi_0$ observed in numerics and supported by filling fraction arguments of MCS theory. We therefore apply one-loop calculations~\cite{azuma2004nonperturbative, Bal:2001cs} to this case, setting the number of fuzzy spheres to $k=2$ and the dimension of the fuzzy spheres to $n=2$.

\begin{figure}[t]
    \centering
    \includegraphics[width= 86mm]{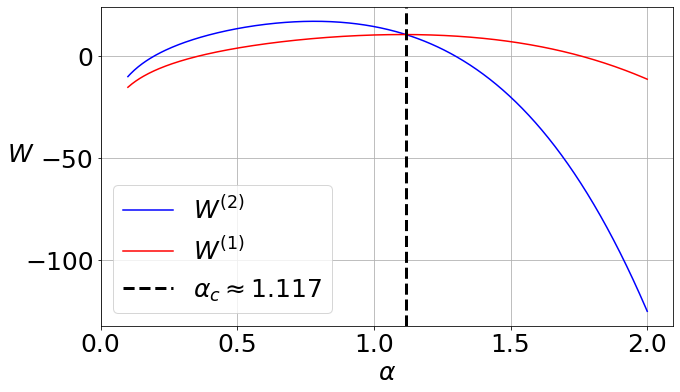}
    \caption{Plot of one-loop corrected free energies $W^{(i)}$ as a function of coupling strength $\alpha$, where $i=1$ represents the configuration with two un-merged fuzzy spheres where $X_\mu = \mathbb{I}\otimes\sigma_\mu$; $i=2$ represents the configuration where $X_\mu$ are $4 \times 4$ irreps of $SU(2)$, corresponding to fuzzy coordinates of a fuzzy sphere with matrix dimension $N=4$. For $\alpha<\alpha_c$ configuration $1$ is preferred, whereas $2$ is preferred for $\alpha>\alpha_c$. At $\alpha=0$ both the free energies diverge to $-\infty$}
    \label{fig:merg}
\end{figure}

\underline{Comparison of free energies:} Let us consider two configurations (1) and (2) - (1) being the case with two $N=2$ fuzzy spheres and (2) being the case with a single $N=4$ fuzzy sphere. In the langauge of Azuma~\emph{et al.}~\cite{azuma2004nonperturbative}, who uses $k$ coincident fuzzy spheres with equal representation sizes $n$, we have $(n,k)=(2,2)$ and $(4,1)$ as the two cases. Here, the total matrix dimension $N=nk=4$ is fixed. The free energy (action) for the classical ground state for case (2) is given by (equation B.12 of ~\cite{azuma2004nonperturbative}):
\begin{equation}
    W^{(2)}_0 = -\frac{1}{24}\alpha^4 N^2(N^2-1) = - 10\alpha^4,
\end{equation}
The one-loop correction to the action is given by (equation B.16 of ~\cite{azuma2004nonperturbative}):
\begin{align}
    W^{(2)}_1 &= \frac{1}{2}\sum^{N-1}_{l=1} (2l+1)\log[N \alpha^2 l(l+1)] \\
    &= 15 \log(\alpha)+\frac{1}{2}\log(8^3 24^548^7) ,
\end{align}
Thus, the total free energy for case (2) is given by:
\begin{equation}
    W^{(2)} = - 10\alpha^4 +  15 \log(\alpha)+\frac{1}{2}\log(8^3 24^548^7),
\end{equation}
Similarly, the free energy of case (1) (from equation B.19 of ~\cite{azuma2004nonperturbative}) is:
\begin{equation}
    W^{(1)}_0 = -\frac{1}{24}\alpha^4 N^2(n^2-1) = -2\alpha^4,
\end{equation}
And the one-loop correction is (from equation B.22 of ~\cite{azuma2004nonperturbative}):
\begin{equation}
    W^{(1)}_1 = \frac{1}{2}k^2\sum^{n-1}_{l=1} (2l+1)\log[N\alpha^2 l(l+1)] = 6\log(8\alpha^2),
\end{equation}
The total free energy for case (1) is then:
\begin{equation}
    W^{(1)} = 6\log(8\alpha^2) - 2\alpha^4,
\end{equation}
If we take the log to be a natural logarithm, then the graphical analysis shown in Fig.~\ref{fig:merg} tells us that for the value of $\alpha_c \approx 1.11736$ there is a transition between cases (1) and (2). Case (1) is more stable for $0<\alpha<\alpha_c$ and case (2) is more stable for $\alpha>\alpha_c$. This is quite similar to numerics for the MCI~\cite{banerjee2024multiplicative}, which suggest merging of the two $N=2$ parent Hamiltonians, within finite $N$ MCS theory, at flux $\phi$ very close to $1$ in units of the flux quantum $\phi_0$. \\

\subsection{Generalizations with anisotropic fuzzification}

A number of generalizations of the above description are possible and we discuss some more immediately obvious examples. Importantly, within this treatment, it is possible to consider distinct topology for each small $N$ Lagrangian, or $k_i \neq k_j$, as depicted in Fig.~\ref{lowsymmdropletarray}. This construction may then be useful in understanding QH systems at dilute filling fractions, an open question of MCS theory raised by Susskind~\cite{susskind2001quantum} and Polychronakos~\cite{polychronakos2001quantum}. For instance, dilute filling fraction might be modeled in part by a small fraction of the $M$ MCS droplets in the array having CS level $k$ non-trivial, and all other MCS droplets individually being topologically trivial. MCS theories within anisotropic fuzzification could be far richer, however, and should be investigated for more general cases of dependence of CS level $k$ for an individual MCS/CSM droplet on its coordinates within the $D$-dimensional array of small $N$ droplets. Additionally, it is important to study the arrays of $M$ MCS droplets, each with $N$ small, as potentially characterizing generalizations of FQH states when the droplets remain distinguishable from one another in some sense, such as in having distinguishable filling fractions. Individual MCS/CSM droplets could also possess gaps to excitations or be gapless. 

 As well, introducing dependence of the positions of the potential well centers on additional parameters and/or lowering the symmetry of the arrangement of droplets, the above formulation could describe a generalized fluid. Related to this point, the electron number for each droplet could also be made dependent on $i$, or $N \rightarrow N_i$, potentially as part of modeling different filling fractions $k_i$.  $N$ or $N_{i}$ could also potentially be made dependent on time $t$ as well. Finally, couplings between droplets could be generalized, taking inspiration in part from various generalizations of CSMs\cite{PhysRevLett.60.635, PhysRevLett.60.639, pasquier2005lecture, Cappelli:2009pn}.

\subsubsection{Relevance of present constructions to spin lattice models and lattice gauge theories}

\begin{figure}[tbp]
    \centering
  \includegraphics[width=\columnwidth]{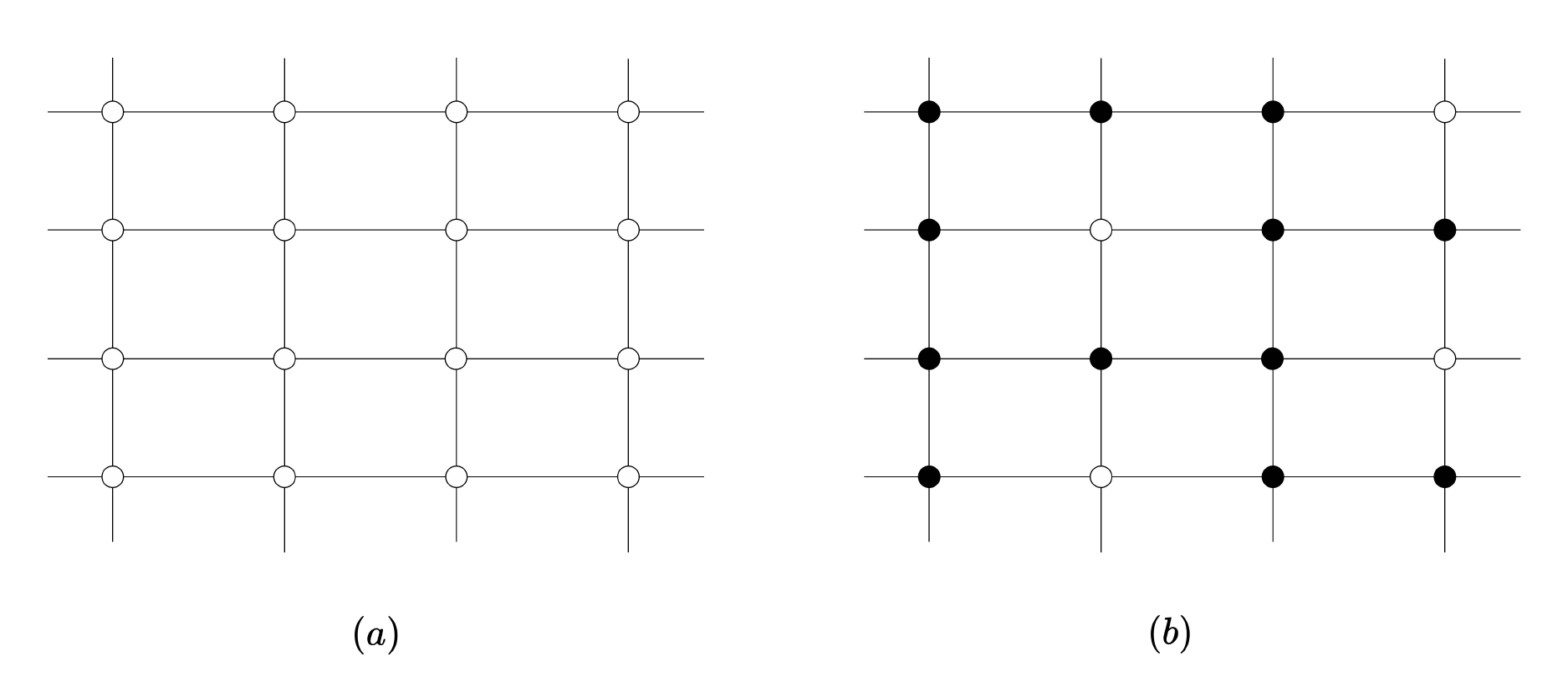}
  \caption{(a) Microscopic field theory of the QSkHE modeled within anisotropic fuzzification as a square lattice, with each vertex generalized to a CSM for a small number of electrons, $N$, corresponding to finite matrix dimension $N$, represented by a ring. Here, each CSM has the same filling fraction $\nu=1/(k+1)$. This uniform lattice can also be describing within the tiling construction. (b) More general microscopic field theory of the QSkHE possible within anisotropic fuzzification, in which case each small $N$ CSM generalizing a vertex of the lattice can possess a distinct filling fraction. The CSM rings are colored white and black in the present schematic to represent two different filling fractions, e.g. white for $\nu=1/(k+1)$, black for $\nu'=1/(k'+1)$, with $k\neq k'$.}
  \label{lowsymmdropletarray}
\end{figure}

\begin{figure}[tbp]
    \centering
  \includegraphics[width=\columnwidth]{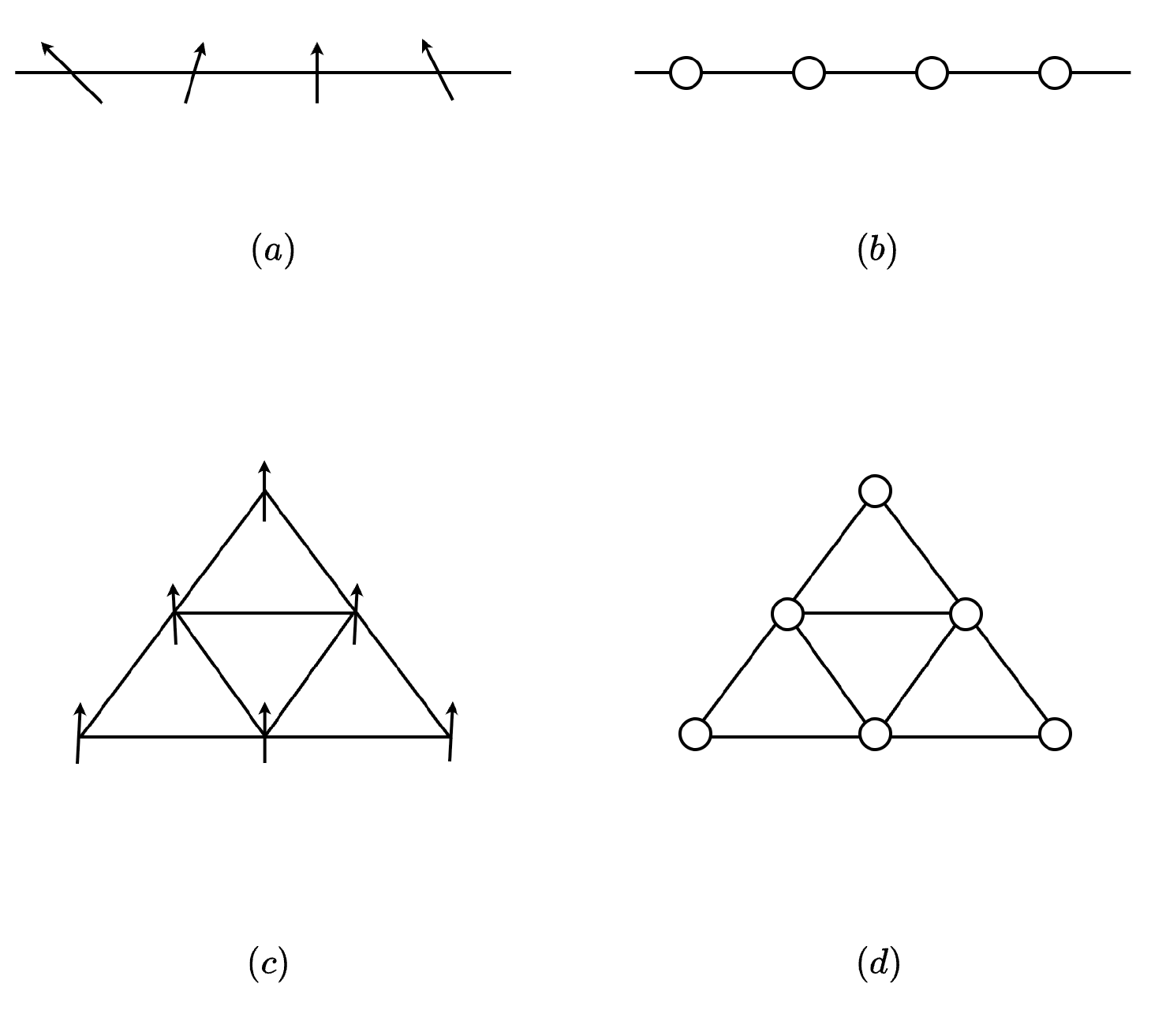}
  \caption{(a) Schematic of a chain of spins, each of spin $S$. (b) Generalization of the spin chain in (a) to a chain of CSMs represented as small rings, each with $2S+1$ spinless electrons. (c) Schematic of a triangular lattice of spins, each of spin $S$. (d) Generalization of the triangular spin lattice according to the present work, as a triangular lattice with vertices generalized to CSMs, each with $2S+1$ spinless electrons.}
  \label{spinlattice_CSM}
\end{figure}

The present work, in formulating signatures of the QSkHE associated with spin DOF, for spin operator matrix dimension $N$, in terms of MCS theory for a finite number $N$ of electrons, motivates generalized treatments of individual spins in various settings. While the present work focuses on modeling spin DOFs of a Bloch Hamiltonian unit cell of matrix dimension $N$, for cases of Bloch Hamiltonians quadratic in second-quantized creation and annihilation operators, strongly-motivating interpretation of these systems previously-treated as effectively non-interacting instead as many-body in some sense, these results are naturally extended to other spin systems. In particular, results in this section on microscopic field theories within anisotropic fuzzification motivate generalized treatments of spin lattice models and related phases of matter, such as quantum spin liquids~\cite{ANDERSON1973153, ANDERSON87, BASKARAN1987973, PhysRevLett.59.2095, PhysRevB.35.8865, PhysRevLett.61.2376, PhysRevB.37.3774, PhysRevLett.66.1773, PhysRevB.40.7387, PhysRevB.44.2664}. A chain of $M$ spins as shown in Fig.~\ref{spinlattice_CSM} (a), each of spin $S$, can be generalized to a chain of $M$ coupled CSM models, each with $N=2S+1$ spinless electrons, as shown in Fig.~\ref{spinlattice_CSM} (b). More generally, a spin $S$ in a lattice gauge theory~\cite{PhysRevLett.61.2376} can be generalized to a CSM with $2S+1$ spinless electrons. A representative example of a triangular spin lattice model is similarly shown in Fig.~\ref{spinlattice_CSM} (c), with its generalization to a lattice of coupled CSM models, each with $2S+1$ spinless electrons, shown in Fig.~\ref{spinlattice_CSM} (d). From this perspective, we believe simpler chiral spin liquids might be well-understood in terms of a lattice of $M$ CSM models, each with the same filling fraction $1/(k+1)$, as shown in Fig.~\ref{lowsymmdropletarray} (a). More complex cases of quantum spin liquids might correspond to more general cases of the individual CSM droplets in the lattice possessing distinct filling fractions, as shown in Fig.~\ref{lowsymmdropletarray} (b).

\section{Discussion and Conclusion}

We introduce microscopic field theories of the quantum skyrmion Hall effect (QSkHE) in this work. That is, we define quantum field theories realizing the QSkHE and related phenomena, explicitly including particles subjected to external fields, which take into account that a spin degree of freedom (DOF) with associated spin operator matrix representation of dimension $N$ encodes a finite number of spatial dimensions for $N$ small, in the regime in which spin has previously been treated as an internal DOF/isospin, effectively only a label of quantum states. We employ machinery of matrix Chern-Simons (MCS) theory~\cite{susskind2001quantum} for a finite number $N$ of spinless electrons in minimal cases~\cite{polychronakos2000noncommutative, polychronakos2001quantum, polychronakos2001quantumcyl}, expressed in terms of matrices of dimension $N$. While motivated in particular by the topological response of the multiplicative Chern insulator (MCI)~\cite{banerjee2024multiplicative}, a canonical example of the multiplicative topological phases~\cite{cook2022multiplicative} which are one of three sets of topological phases within the framework of the QSkHE~\cite{qskhe, patil2024effective}, we find the QSkHE is well-supported on first principles by MCS theory. However, we also introduce some key generalizations of MCS to capture additional features of the QSkHE observed in related works, some of the most essential being Ay~\emph{et al.}~\cite{ay2024signatures},  Banerjee and Cook~\cite{banerjee2024multiplicative}, and Patil~\emph{et al.}~\cite{patil2024effective}, which are however also motivated on first principles. 

We first study a single MCS theory for a finite number of electrons $N$, with $N$ small, such as $N=4$. We employ such a small $N$ MCS/CSM theory for a quantum Hall (QH) droplet with $N$ spinless electrons, or MCS/CSM droplet, to model a single unit cell of a system realizing phenomena of the QSkHE, such as a single quantum spin $S$ of multiplicity $N=2S+1$, or a single unit cell of a crystalline system with an $N$-fold DOF for the unit cell. As part of this analysis, we derive quantization of the topological invariant $\text{Tr}[C]$ introduced to characterize the QSkHE~\cite{banerjee2024multiplicative, patil2024effective}, which defines topological artifacts of the spin operator algebra projected to an occupied subspace. To do so, we consider a single quantum Hall droplet with a finite number $N$ of electrons within MCS theory, defined over a fuzzy two-sphere, or matrix approximation of the two-sphere. We apply primarily the procedure developed by Polychronakos to quantize the filling fraction $\nu$ of the droplet, via mapping to a Calogero-Sutherland model (CSM) for $N$ electrons, a generalized harmonic oscillator in one spatial dimension, with an additional interaction term yielding fractional exclusion statistics. We encode deformations of the fuzzy two-sphere over which the MCS theory is defined due to partial filling of the sphere by converting the Poisson bracket Lie derivative to a quantum counterpart Lie derivative, a Lie derivative of the deformed fuzzy space, which includes a term containing operator $C$ of the invariant $\text{Tr}[C]$. A term for insertion of two space-like Wilson lines in the resultant Gauss law, previously identified as traceless by Polychronakos, instead generalizes to yield fusion rules for charged quasi-hole/particle excitations of the theory, corresponding to insertion instead of three space-like Wilson lines. The trace of the term then more generally is non-zero and quantizes to some multiple of $k+1$, the Chern-Simons level of the finite $N$ MCS theory. Given the possibly quite rich deformations from fuzzy sphere geometry, fusion in the MCS/CSM Gauss law is simply identified in the present work as the origin of non-trivial and quantized $\text{Tr}[C]$, while the specific fusion rules will be studied in detail in future work. In the process of quantizing $\text{Tr}[C]$, we identify quantum skyrmions of the QSkHE as low-lying charged excitations of the finite $N$ MCS theory/CSM, which are their quasi-hole and quasi-particle excitations.

Building on these results, we study particular features of the MCI topological response. First, we explore mechanisms for realizing the $1/3$ contribution observed for $\text{Tr}[C]$ when the MCI is subjected to time-reversal-symmetric flux insertion, a signature of individual unit cells of the MCI lattice in real-space. Employing expressions for filling fractions within MCS theory with a finite number $N$ of electrons, we model the unit cell as a pair of QH droplets, each described by an $N=2$ MCS/CSM theory. This pair of droplets models a single unit cell of the MCI in real-space. As the topological response of the MCI of interest is observed for the case of one parent Hamiltonian with Chern number $+1$ and the other with Chern number $-1$, corresponding to Chern-Simons level in Susskind's unbounded MCS theory of $k=1$, we consider these two $N=2$ droplets each at filling fraction $1/2$ in the counterpart finite $N$ MCS theory. We model the flux insertion process as breaking of the Kronecker product structure, yielding an $N=4$ MCS droplet at filling fraction $1/3$. We find the pair of $N=2$ droplets at filling fraction $1/2$, each contributing two electrons and a single bare flux, have the same number of electrons and bare flux quanta as the single $N=4$ droplet at filling fraction $1/3$ (four electrons and two bare flux) in the ground state. Taking into account that the two $N=2$ droplets merge while inserting a single flux quantum through the MCS/CSM droplet then corresponds to creation of an additional quasi-hole excitation of charge $1/3$ in the simplest case, consistent with the observed $1/3$ contribution to $\text{Tr}[C]$. We furthermore note that, for the Laughlin states currently with the framework of MCS theory, these small $N$ MCS droplets at filling fraction $1/(k+1)$ may be identified as possessing $k$ bare flux quanta in the ground state. We then identify the QH droplet of the MCS theory itself, or MCS/CSM droplet, as a generalization of a point charge. Upon integrating out the internal DOFs of the droplet to arrive at a point charge, we identify the point charge and $k$ flux quanta as a Jain composite particle for the Laughlin state at filling fraction $1/k$~\cite{jain1994}. We therefore propose the small $N$ MCS QH droplet as a generalization of the Jain composite particle, currently only for filling fractions of the Laughlin states. This motivates future work to extend MCS theory to more complex filling fractions associated with the rest of the hierarchy of the fractional quantum Hall (FQH) states. We will discuss one possibility for achieving such generalizations later in this discussion section.

Employing a tiling construction of Polychronakos previously-used to construct a $1+1$ dimensional, commutative, U(N) non-Abelian Yang-Mills action from the finite $N$ MCS theory, we then construct microscopic field theories of the QSkHE at the level of extra fuzzy dimensions~\cite{patil2024effective}. We instead construct a $2+2+1$ dimensional commutative, non-Abelian Yang-Mills action retaining the two spatial dimensions encoded by the finite $N$ MCS theory in addition to two Cartesian space coordinates introduced by the tiling procedure, and take into account generalization from U(N) structure due to partial filling of the finite $N$ MCS droplet. Through this process, the boundary term of the finite $N$ MCS theory shifting the filling fraction from $1/k$ for the unbounded theory to $1/(k+1)$ for the bounded theory becomes a source term in the $2+2+1$ dimensional Lagrangian, and the filling fraction of the $2+2+1$ dimensional theory is then $1/k$. This is consistent with the topological response of the MCI: the entire system exhibits a generalized $4 \pi$ Aharonov-Bohm effect, which can be identified with filling fraction $1/2$, in combination with a $1/3$ contribution to the topological invariant $\text{Tr}[C]$ for two individual unit cells of the MCI lattice model in real-space. These results furthermore strongly support generalizing interpretation of the spin $S$ DOF associated with the unit cell defined by a general Bloch Hamiltonian of matrix dimension $2S+1$, the spin multiplicity, to an effective MCS theory/CSM with $2S+1$ spinless electrons within currently available machinery. We find such a generalization is relevant even for Bloch Hamiltonians, which are quadratic in second-quantized creation and annihilation operators. In this sense, even these relatively simple Hamiltonians can actually capture at least some many-body physics, and therefore realize signatures of generalized FQH states within MCS theory. We finally note that the tiling construction is invaluable for identifying certain terms and processes of commutative, non-Abelian quantum field theories with counterparts in finite $N$ MCS theory and the QSkHE. As well as the boundary term of finite $N$ MCS theory being a source term in the commutative, non-Abelian gauge theory, an applied electric field within the finite $N$ MCS theory is equivalent to an effective Zeeman field term in the counterpart commutative, non-Abelian gauge theory. Similarly, flux insertion through the plane of the commutative, non-Abelian gauge theory may be identified with flux insertion for the finite $N$ MCS theory, which is more easily visualized as flux insertion through the CSM ring. These constructions strongly suggest that first experimental observation of signatures of the QSkHE itself, rather than lattice counterparts, may come from non-Abelian FQH states. This strongly motivates careful examination in particular of quantum Hall bilayers~\cite{girvin1996multicomponent}, as well as development of experimental methods for directly measuring the quantum skyrmion charge encoded by the topological invariant $\text{Tr}[C]$.

Finally, we construct microscopic field theories of the QSkHE within the framework of anisotropic fuzzification introduced in Patil~\emph{et al.}~\cite{patil2024effective}. To do so, we extend methods for constructing Lagrangians for multiple fuzzy spheres to Lagrangians for multiple finite $N$ MCS droplets defining deformations of fuzzy spheres. Within this framework, we may define $D$-dimensional arrays consisting of $M$ MCS/CSM droplets, each with its own, potentially unique, number of electrons and filling fraction, as well as coupling between these droplets. Defining specific inter-droplet couplings is a rich area of future research. While we focus on crystalline arrays of $M$ droplets, this construction is readily generalized to describe non-crystalline arrangements of the $M$ droplets that could describe generalized fluids or amorphous solids, as some examples. It may also be important and fruitful to take into account possible vibrational modes of the $M$ MCS droplets, possibly from the perspective of their counterpart CSMs. 

Within this formalism, it is possible to consider arrays in which only a small number of droplets are at non-trivial Chern-Simons level, which may be useful in applying MCS theory to study QH systems at dilute filling fraction, an open problem of MCS theory. We expect this framework to also be useful in study of disorder. Lagrangians for multiple fuzzy spheres have furthermore previously been applied to study merging of multiple fuzzy spheres into a single fuzzy sphere. We apply some of these results to model the merging of the two $N=2$ MCS/CSM droplets, modeling a single unit cell of the MCI in real-space, into a single $N=4$ MCS/CSM droplet as a consequence of local flux insertion. Although deformations due to partial filling of fuzzy spheres are not taken into account, the $N=4$ configuration becomes energetically favorable in comparison to the configuration with two $N=2$ spheres, for coupling strength of the interaction term of approximately $1$. This appears to be consistent with emergence of the $1/3$ contribution to $\text{Tr}[C]$ for flux $\phi$ of close to one flux quantum as part of the topological response of the MCI, as the $1/3$ contribution corresponds to merging of two $N=2$ droplets into a single $N=4$ droplet as part of flux insertion according to MCS theory, which then also produces a quantum skyrmion of charge $1/3$.

The present work, particularly the results on anisotropic fuzzification, suggest MCS theories at finite matrix dimension $N$ are generally composite outside of simple cases, consisting of multiple coupled component MCS droplets, each with $n_i$ electrons, with $\sum_i n_i = N$. Future work will explore modeling of these composite objects in terms of nested fuzzy spaces for individual unit cells ($N$ small), in analogy to construction of higher-dimensional fuzzy spheres as nested fuzzy two-spheres~\cite{hasebe2004dimensional}, similarly to the nesting of fuzzy spheres discussed here for modeling multiple, coupled unit cells within anisotropic fuzzification. Interpretation of the MCI topological response within MCS theory strengthens the case for pairing between MCS droplets induced by flux insertion. The entanglement associated with this pairing/merging might itself furthermore be characterized, possibly in terms of quantum skyrmions in analogy to past work on entanglement skyrmions~\cite{PhysRevB.78.195327}.

An important step in extending MCS theory further will be to realize more general filling fractions than those of the Laughlin states. Anisotropic fuzzification may be one path to realizing such more general filling fractions $\nu = p/q$, with $p$, $q$ each integer, as it permits scenarios, for instance, where each level of the first fuzzy two-sphere in the nesting construction of the fuzzy four-sphere effectively can hold $N>1$ electrons of the second fuzzy sphere in the nesting construction. More general filling fractions might correspond to e.g., every $j$\textsuperscript{th} orbital of the first fuzzy two-sphere occupied effectively by $N>1$ electrons of the second fuzzy sphere. We will explore these and other possibilities for realizing more general filling fractions in future work.

Generalization of a single spin $S$ of multiplicity $2S+1$ to a MCS droplet/CSM with $2S+1$ electrons, and generalization of the MCS droplet/CSM to include fusion---as is actually consistent both with the differential geometry of partially-filled, fuzzy coset spaces and expected given the quasi-hole/particle excitations possible in these systems (the quantum skyrmions of the QSkHE)---necessitates careful examination of related topics, particularly given the many deep connections between the CSM and other topics in theoretical physics. As an example, random matrix theory, while typically treated at $N \rightarrow \infty$~\cite{PhysRevB.52.8729, altland1997nonstandard}, should be studied in greater depth at finite---and even small---$N$. Existing works for finite $N$ should be examined for potential dependence on spherical symmetry which should be generalized due to deformations associated with partial filling~\cite{RDelannay_2000}. Even more importantly, perhaps, the present work motivates modeling an individual spin $S$, of multiplicity $2S+1$, in terms of a MCS theory/CSM with $2S+1$ spinless electrons. This is naturally extended to modeling lattices of spins, each of spin $S$, as lattices of MCS/CSM droplets, each droplet with $2S+1$ spinless electrons. Quantum dimer models and more general lattice gauge theories can then also be generalized in this manner. Such generalizations will likely yield valuable insights into the nature of myriad strongly-correlated phases of matter of spin systems, such as quantum spin liquids. Given signatures of pairing between finite $N$ MCS/CSM droplets, these results may also be relevant to understanding of some forms of unconventional superconductivity and related strongly-correlated states, such as non-Fermi liquids \cite{PhysRevB.14.1165, PhysRevB.48.7183, RevModPhys.79.1015, RevModPhys.73.797, PhysRevB.78.035103}, given that the finite $N$ MCS droplet and its quantum skyrmion excitations generalize the concept of a quasi-particle.

Finally, we note that identification of fusion in finite $N$ MCS theory strongly motivates investigation of the QSkHE from a quantum information/computing perspective. The fuzzy two-sphere of finite matrix dimension $N$ has already been identified as encoding some number of qubits, effectively modeling a quantum computer~\cite{zizzi20142}. In this work, we identify deformations and fusion associated with partial-filling of finite $N$ fuzzy spheres, and we identify these partially-filled fuzzy spaces with potentially even an individual spin $S$ of multiplicity $N = 2S+1$. The phenomena reported here are therefore promising in developing more robust quantum computation schemes, and are expected to yield generalized quantum gate operations~\cite{nayak2008}. The possibility of fusion and arranging a single spin in topologically-distinct states characterized by $\text{Tr}[C]$ suggests even measurement in quantum mechanics can be re-examined and potentially better understood/controlled~\cite{zizzi20142}. A natural starting point in identifying the fusion rules is the effective two-qubit states reported for the multiplicative Kitaev chain~\cite{pal_mkc}, which implicitly define potential gate operations~\cite{kitaev2003fault}.

\section{Appendix: Acronyms and Conventions}
\begin{itemize}
    \item Aharonov-Bohm (AB)
    \item Calogero-Sutherland Model (CSM)
    \item Chern-Simons theory (CS)
    \item Degree of freedom (DOF)
    \item Effective field theory (EFT)
    \item Equation of motion (EOM)
    \item Fractional quantum Hall (FQH) 
    \item Fractional quantum Hall effect (FQHE)
    \item Landau level (LL)
    \item Lowest Landau level (LLL)
    \item Matrix Chern-Simons theory (MCS)
    \item Open boundary conditions (OBCs)
    \item Periodic boundary conditions (PBCs)
    \item Quantum Hall (QH)
    \item Quantum Hall effect (QHE)
    \item Quantum Hall ferromagnet (QHFM)
    \item Quantum skyrmion Hall effect (QSkHE)
\end{itemize}
\bibliography{ref}

\end{document}